\documentclass[twocolumn,longbib]{aastex701}

\usepackage{makecell,CJK}
\usepackage{multirow}

\newcommand{\jwst}{JWST}
\newcommand{\hst}{HST}

\definecolor{purp}{rgb}{0.62, 0.0, 1.0}

\newcommand{\auburn}{Department of Physics, Auburn University, Edmund C.\ Leach Science Center, Auburn, 36849, AL, USA}
\newcommand{\esaneocc}{ESA PDO NEO Coordination Centre, Largo Galileo Galilei, 1, I-00044 Frascati (RM), Italy}
\newcommand{\jpl}{Jet Propulsion Laboratory, California Institute of Technology, 4800 Oak Grove Dr., Pasadena, CA 91109, USA}
\newcommand{\lowell}{Lowell Observatory, 1400 W.\ Mars Hill Rd., Flagstaff, AZ 86001, USA}
\newcommand{\planetaryscienceinst}{Planetary Science Institute, 1700 East Fort Lowell Rd., Suite 106, Tucson, AZ 85719, USA}
\newcommand{\umd}{Department of Astronomy, University of Maryland, 4296 Stadium Dr., College Park, MD 20742, USA}
\newcommand{\villanova}{Department of Astrophysics and Planetary Science, Villanova University, Villanova, PA, 19085, USA}
\newcommand{\carnegie}{Earth and Planets Laboratory, Carnegie Institution for Science, 5241 Broad Branch Road NW, Washington, DC 20015, USA}
\newcommand{\uwdirac}{Department of Astronomy \& the DiRAC Institute, University of Washington, 3910 15th Ave NE, Seattle, WA 98195, USA}
\newcommand{\lincc}{LSST Interdisciplinary Network for Collaboration and Computing, 933 N.\ Cherry Avenue, Tucson, AZ 85721, USA}
\newcommand{\nau}{Department of Astronomy \& Planetary Science, Northern Arizona University, P.O.\ Box 6010, Flagstaff, AZ 86011, USA}
\newcommand{\rawdata}{Raw Data Speaks Initiative, USA}
\newcommand{\edinburgh}{Institute for Astronomy, University of Edinburgh, Royal Observatory, Edinburgh EH9 3HJ, UK}

\providecommand\aap{Astron.\ Astrophys.} % Astron and Astrophys
\providecommand\apjl{Astrophys.\ J.} % Astrophysical Journal, Letters
\providecommand\apjs{Astrophys.\ J.\ Suppl.} % Astrophysical Journal, Supplement
\received{December 8, 2025}
\accepted{April 23, 2026}
\submitjournal{PSJ}

\shorttitle{Volatile Composition of 133P and Other MBCs}
\shortauthors{Hsieh et al.}

\hypersetup{pdfauthor={Name}}
\begin{document}
\begin{CJK*}{UTF8}{gbsn}

\title{Characterization of the Volatile Properties of 133P/Elst-Pizarro and Other Main-Belt Comets with \jwst{} and Ground-Based Observations}

\correspondingauthor{Henry Hsieh}
\email{hhsieh@psi.edu}

\author[0000-0001-7225-9271]{Henry H.\ Hsieh}
\affiliation{\planetaryscienceinst}
\email{hhsieh@psi.edu}

\author[0000-0003-2152-6987]{John W.\ Noonan}
\affiliation{\auburn}
\email{noonan@auburn.edu}

\author[0000-0002-6702-7676]{Michael S.\ P.\ Kelley}
\affiliation{\umd}
\email{msk@astro.umd.edu}

\author[0000-0002-2668-7248]{Dennis Bodewits}
\affiliation{\auburn}
\email{dzb0059@auburn.edu}

\author[0000-0002-5736-1857]{Jana Pittichov\'a}
\affiliation{\jpl}
\email{jana.chesley@jpl.nasa.gov}

\author[0000-0002-1506-4248]{Audrey Thirouin}
\affiliation{\lowell}
\email{thirouin@lowell.edu}

\author[0000-0001-7895-8209]{Marco Micheli}
\affiliation{\esaneocc}
\email{marco.micheli@esa.int}

\author[0000-0003-3145-8682]{Scott S.\ Sheppard}
\affiliation{\carnegie}
\email{ssheppard@carnegiescience.edu}

\author[0000-0001-7335-1715]{Colin O.\ Chandler}
\affiliation{\uwdirac}
\affiliation{\lincc}
\affiliation{\nau}
\affiliation{\rawdata}
\email{coc123@uw.edu}

\author[0000-0003-1008-7499]{Theodore Kareta}
\affiliation{\lowell}
\affiliation{\villanova}
\email{theodore.kareta@villanova.edu}

\author[0000-0001-9328-2905]{Colin Snodgrass}
\affiliation{\edinburgh}
\email{csn@roe.ac.uk}

\author[0009-0007-5946-8731]{Richard E. Cannon}
\affiliation{\edinburgh}
\email{richard.cannon@ed.ac.uk}

\author[0000-0002-8137-5132]{Brian P. Murphy}
\affiliation{\edinburgh}
\email{brian.murphy@ed.ac.uk}

\begin{abstract}
We report results from an analysis of the volatile composition and evolution of main-belt comet (MBC) 133P/Elst-Pizarro using JWST NIRSpec and NIRCam observations and ground-based observations during its 2024 active apparition, and also assess the body of \jwst{} MBC observations acquired to date.  Using NIRSpec, we measure water vapor outgassing rates at two points in 133P's orbit, finding $Q_{\rm H_2O}=(1.9\pm0.6)\times10^{25}$~molecules~s$^{-1}$ on UT 2024 June 12 (at a true anomaly of $\nu=8^{\circ}$ and heliocentric distance of $r_h=2.674$~au), and $Q_{\rm H_2O}=(1.4\pm0.4)\times10^{25}$~molecules~s$^{-1}$ on UT 2024 October 14 (at $\nu=37.4^{\circ}$ and $r_h=2.747$~au). These measurements nominally represent a decline of $\sim$25\% in $Q_{\rm H_2O}$ between the visits, although they are also consistent with no change within uncertainties.  We do not detect CO, CO$_2$, or CH$_3$OH, placing 133P's hypervolatile depletion ($Q_{\rm CO_2}/Q_{\rm H_2O}<0.009$) at a similar level found for previously observed MBCs.  We find $\log(Af\rho/Q_{\rm H_2O})$ values for the three MBCs for which water vapor outgassing has been successfully detected that are consistent within uncertainties with an average value of $\log(Af\rho/Q_{\rm H_2O})=-24.6\pm0.2$.
Lastly, we find no clear correlations of water production rates with nucleus size, semimajor axis, or heliocentric distance among MBCs observed by JWST so far, but would particularly encourage future JWST observations of additional MBCs interior to the 5A:2J MMR with Jupiter and at high inclinations, as well as multiple observations of MBCs during single active apparitions to further investigate areas of interest identified from the current sample of JWST-observed MBCs.
\end{abstract}

\keywords{Main belt comets --- Comets --- Main belt asteroids}

\section{Introduction\label{section:intro}}
\setcounter{footnote}{0}

\subsection{Background\label{section:background}}

Active asteroids are small solar system bodies with asteroid-like orbits \citep[typically parameterized by the Tisserand parameter, $T_J$, where objects with $T_J>3$ are considered to be dynamically asteroidal;][]{vaghi1973_tisserand} that exhibit comet-like dust emission \citep{jewitt2015_actvasts_ast4,jewitt2024_continuum_comets3}.
They are comprised of main-belt comets \citep[MBCs;][]{hsieh2006_mbcs}, whose activity has been determined to be driven (at least partially) by the sublimation of volatile ice, and disrupted asteroids \citep[e.g.,][]{hsieh2012_scheila}, whose activity is due to non-sublimation-related mechanisms such as impacts or rotational destabilization events.  The dynamically stable nature of most of the known MBCs suggest that they either formed in situ or were at least delivered to the asteroid belt at very early times.  Then considering their apparent volatile content as well, MBCs are considered potentially extremely interesting probes of the abundance, distribution, and composition of ice in the inner solar system, and thus are considered interesting subjects for studies of astrobiology, primordial water delivery to the terrestrial planets, and solar system formation and evolution in general \citep[e.g.,][]{hsieh2014_mbcsiausproc}.

Until recently, the sublimation-driven nature of MBC activity was inferred indirectly from dust modeling analyses indicating prolonged durations of dust emission events, and observations of recurrent activity near perihelion with intervening periods of inactivity away from perihelion.  Such behavior is easily explained if the observed activity is driven by sublimation, but difficult to explain as the natural consequence of other mechanisms like impacts or rotational disruptions \citep[e.g.,][]{jewitt2024_continuum_comets3}.  Despite many attempts using some of the largest and most capable telescopes in the world and in space, however, direct detection of sublimation products from active MBCs remained elusive for many years \citep[see references in][]{snodgrass2017_mbcs}.  This situation changed with the first direct detection of water vapor outgassing from MBC 238P/Read by \citet{kelley2023_jwst238p}, who used \jwst{}'s NIRSpec instrument to directly measure a water production rate of $Q_{\rm H_2O}=(9.9\pm1.0)\times10^{24}$~molecules~s$^{-1}$.

Notably, \citet{kelley2023_jwst238p} also found extreme hypervolatile depletion in 238P, finding $Q_{\rm CO_2}/Q_{\rm H_2O}<0.007$, which was about an order of magnitude lower than similar previous spectroscopic measurements of other comets at similar heliocentric distances, and a factor of three lower than any other previous comet measurement overall.  This result was consistent with thermal modeling results suggesting that long-lived dynamically stable objects in the asteroid belt could be strongly depleted in all volatile ices other than water \citep{prialnik2009_mbaice}.  This is significant because many previous efforts to detect sublimation products from MBCs from ground-based facilities targeted proxy species for water production, primarily CN, and then inferred equivalent water production rate upper limits based on assumptions of similar compositional ratios in MBCs as in other comets \citep[see][]{snodgrass2017_mbcs}.  The confirmation that volatile species other than water could be highly depleted in MBCs has since shifted the focus of ongoing characterization of the volatile composition of MBCs almost entirely to \jwst{}, given its unprecedented sensitivity and ability to directly search for water.

Shortly after the initial detection of water vapor outgassing from 238P, water vapor outgassing was detected from a second MBC, 358P/PANSTARRS, by \citet{hsieh2025_358p}, who found even stronger outgassing at the level of $Q_{\rm H_2O}=(5.0\pm0.2)\times10^{25}$~molecules~s$^{-1}$ and similarly strong hypervolatile depletion ($Q_{\rm CO_2}/Q_{\rm H_2O}<0.002$).  Motivated by these two successful detections, we used \jwst{} to observe the first known and currently best-characterized MBC, 133P/Elst-Pizarro, during its 2024 perihelion passage.  In contrast to 238P and 358P, which \jwst{} only observed once during their active phases, we observed 133P at two different heliocentric distances and orbit positions in 2024. This enables us to characterize for the first time the evolution of a MBC's active behavior over the course of the same active apparition.  In this manuscript, we report and discuss the results of these observations as well as results from a concurrent ground-based observing campaign conducted in support of those observations.

\setlength{\tabcolsep}{4pt}
\setlength{\extrarowheight}{0em}
\begin{table*}[htb]
\caption{\jwst{} 133P Observations$^a$}
\centering
\smallskip
\footnotesize
\begin{tabular}{clccrrrrccrrr}
\hline\hline
\multicolumn{1}{c}{Target}
 & \multicolumn{1}{c}{UT Date}
 & \multicolumn{1}{c}{UT Time}
 & \multicolumn{1}{c}{Instrument}
 & \multicolumn{1}{c}{$\nu$$^b$}
 & \multicolumn{1}{c}{$r_h$$^c$}
 & \multicolumn{1}{c}{$\Delta_{\oplus}$$^d$}
 & \multicolumn{1}{c}{$\Delta_{obs}$$^e$}
 & \multicolumn{1}{c}{$\alpha_{\oplus}$$^f$}
 & \multicolumn{1}{c}{$\alpha_{\rm obs}$$^g$}
 & \multicolumn{1}{c}{PA$_{-\odot}$$^h$}
 & \multicolumn{1}{c}{PA$_{-v}$$^i$}
 & \multicolumn{1}{c}{$\Delta t_q$$^j$}
 \\
\hline
133P & 2024 Jun 12 & 13:18:36 - 15:42:03 & NIRSpec &  8.0 & 2.674 & 1.980 & 1.973 & 18.5 & 18.8 & 251.0 & 253.6 & $+$33 \\ % obs=2460473.8 q=2460440.8
133P & 2024 Oct 14 & 07:50:00 - 08:33:50 & NIRCam  & 37.4 & 2.747 & 2.268 & 2.269 & 20.2 & 20.4 &  74.8 & 256.1 & $+$157 \\ % obs=2460597.8 q=2460440.8
133P & 2024 Oct 14 & 09:03:53 - 11:37:33 & NIRSpec & 37.4 & 2.747 & 2.270 & 2.271 & 20.2 & 20.4 &  74.8 & 256.1 & $+$157 \\ % obs=2460597.8 q=2460440.8
133P & 2024 Oct 28 & 11:49:49 - 12:15:46 & NIRCam  & 40.6 & 2.760 & 2.466 & 2.470 & 20.9 & 21.2 &  74.0 & 255.3 & $+$171 \\ % obs=2460611.8 q=2460440.8
\hline
\hline
\multicolumn{13}{l}{$^a$ Observing geometry parameters from JPL Horizons (Solution \#77) \citep{giorgini1996_horizons}} \\
\multicolumn{13}{l}{$^b$ True anomaly, in degrees} \\
\multicolumn{13}{l}{$^c$ Heliocentric distance, in au} \\
\multicolumn{13}{l}{$^d$ Geocentric distance, in au} \\
\multicolumn{13}{l}{$^e$ \jwst{}-centric distance, in au} \\
\multicolumn{13}{l}{$^f$ Solar phase angle (Sun-target-Earth), in degrees} \\
\multicolumn{13}{l}{$^g$ Solar phase angle (Sun-target-\jwst{}), in degrees} \\
\multicolumn{13}{l}{$^h$ Position angle of the anti-Solar vector as projected on the sky, in degrees East of North} \\
\multicolumn{13}{l}{$^i$ Position angle of the negative heliocentric velocity vector as projected on the sky, in degrees East of North} \\
\multicolumn{13}{l}{$^j$ Time relative to perihelion (positive values indicating time after perihelion), in days.} \\
\end{tabular}
\label{table:jwst_mbc_observations}
\end{table*}

\subsection{133P/Elst-Pizarro\label{section:background_133p}}

Discovered as a comet in 1996 \citep{elst1996_133p}, 133P was the first comet-like object observed within the main asteroid belt \citep{hsieh2004_133p}. Its reactivation in 2002 provided strong evidence that the observed activity is driven by the periodic sublimation of volatile ices rather than by stochastic processes such as impacts. Recurrent activity has since been confirmed during five consecutive perihelion passages between 1996 and 2019 \citep[][and unpublished data from 2019 obtained by the authors of this work]{elst1996_133p,hsieh2004_133p,hsieh2010_133p,hsieh2013_133p,jewitt2007_133p,jewitt2014_133p}. Its low eccentricity and inclination suggest that 133P is unlikely to be a Jupiter-family comet interloper \citep{hsieh2016_tisserand}.

Observations from the {\it Spitzer Space Telescope} indicate an exceptionally low mean Bond albedo of 0.024, while data from {\it Spitzer} and the Hubble Space Telescope (\hst{}) suggest a nucleus diameter of approximately 4 km \citep{hsieh2009_albedos, jewitt2007_133p, yu2020_elstpizarro}, making 133P among the largest known main-belt comets. 
Lightcurve analyses show a peak-to-trough photometric range of $\Delta m\sim0.4$~mag, corresponding to projected dimensions of approximately 4.6~km $\times$ 3.2~km with an axis ratio of $a/b=1.45$, and a rotation period of $P_{\rm rot}=(3.471\pm0.001)$~hr \citep{hsieh2004_133p,hsieh2009_albedos,jewitt2014_133p}.

When active, 133P has typically been observed having no resolved coma but possessing a narrow dust tail that is observable for months. Numerical dust modeling analyses have indicated that these observations are consistent with sustained, low-level dust emission persisting for periods of months \citep{hsieh2004_133p, jewitt2014_133p}. 
A search for CN emission during 133P's 2007 active apparition using the European Southern Observatory (ESO) Very Large Telescope (VLT) resulted in a non-detection, indicating that if outgassing was present, it was below the detection limit of the observations \citep{licandro2011_133p176p}.  This result was consistent with other attempts to directly detect outgassing from MBCs using ground-based facilities at the time \citep{snodgrass2017_mbcs}.

\section{Observations\label{section:observations}}

\jwst{} \citep{gardner2023_jwst} observations of 133P were obtained by NIRCam
\citep{rieke2023_nircam}
on UT 2024 October 14 and UT 2024 October 28, and by NIRSpec 
\citep{jakobsen2022_nirspec,boker2023_nirspec}
on UT 2024 June 12 and UT 2024 October 14 as part of \jwst{} General Observer (GO) programs GO 4250\footnote{\url{https://www.stsci.edu/jwst/science-execution/program-information?id=4250}} in Cycle 2 and GO 5551\footnote{\url{https://www.stsci.edu/jwst/science-execution/program-information?id=5551}} in Cycle 3.  Observational circumstances of these observations are listed in Table~\ref{table:jwst_mbc_observations}.
Due to technical issues, the GO 4250 NIRCam observation of 133P could not be executed during the available visibility window in Cycle 2, and so was instead executed during Cycle 3 along with another observation of 133P that was already planned as part of GO 5551, hence the close temporal proximity of those observations.

\setlength{\tabcolsep}{10pt}
\setlength{\extrarowheight}{0em}
\begin{table*}[htb]
\caption{Ground-Based Observing Instrumentation Characteristics}
\centering
\smallskip
\footnotesize
\begin{tabular}{lcccc}
\hline\hline
\multicolumn{1}{c}{Telescope$^a$}
 & \multicolumn{1}{c}{Instrument}
 & \multicolumn{1}{c}{FOV$^b$}
 & \multicolumn{1}{c}{Pixel Scale$^c$}
 & \multicolumn{1}{c}{Binning}
 \\[2pt]
\hline
Magellan & IMACS   & $15\farcm4\times15\farcm4$ & 0.20  & $1\times1$ \\
Gemini-N & GMOS-N  & $5\farcm5\times5\farcm5$   & 0.16  & $2\times2$ \\
Gemini-S & GMOS-S  & $5\farcm5\times5\farcm5$   & 0.16  & $2\times2$ \\
NTT      & EFOSC2  & $3\farcm9\times3\farcm9$   & 0.24  & $2\times2$ \\
Palomar  & WaSP    & $18\farcm4\times18\farcm5$ & 0.175 & $1\times1$ \\
LDT      & LMI    & $12\farcm3\times12\farcm3$ & 0.24 & $2\times2$ \\
\hline
\hline
\multicolumn{5}{l}{$^a$ Magellan: Magellan Baade telescope; Gemini-N: Gemini North} \\
\multicolumn{5}{l}{$~~~$ telescope; Gemini-S: Gemini South telescope; NTT: New Technology} \\
\multicolumn{5}{l}{$~~~$ Technology Telescope; Palomar: Palomar Hale Telescope; LDT:} \\
\multicolumn{5}{l}{$~~~$ Lowell Discovery Telescope (in 2$\times$2 binning mode)} \\
\multicolumn{5}{l}{$^b$ Field of view dimensions.} \\
\multicolumn{5}{l}{$^c$ Pixel scale in arcsec~pixel$^{-1}$ at specified binning.} \\
\end{tabular}
\label{table:instrumentation}
\end{table*}

NIRCam observations of 133P were acquired simultaneously in the F200W and F277W broadband filters, using separate detectors fed by a dichroic. Each detector is $2040\times2048$ pixels in size, but with pixel scales of $0\farcs0313$~pixel$^{-1}$ and $0\farcs0630$~pixel$^{-1}$ for the F200W and F277W images, respectively. These different scales provide angular fields of view of $63\farcs9\times64\farcs1$ (F200W images) and $128\farcs5\times129\farcs0$ (F277W images), respectively.  When observing a Solar spectrum \citep{wilmer2018_solarspectrum}, the F200W and F277W filters (spanning wavelengths between 1.725~$\mu$m to 2.260~$\mu$m, and 2.367~$\mu$m to 3.220~$\mu$m, respectively) are characterized by effective wavelengths of 1.97~$\mu$m and 2.74~$\mu$m, respectively \citep{kelley2023_jwst238p}. The 4-point INTRAMODULEBOX dither pattern was used to enable mitigation of detector artifacts and cosmic rays, with one exposure at each dither point.  The SHALLOW4 readout pattern was used for each dither exposure to provide a total of 1031~s per filter.

Both our NIRSpec observations of 133P and an off-source background field were acquired using the NIRSpec integral field unit (IFU) to obtain spatially resolved imaging spectroscopy for a $3''\times3''$ field of view. The IFU observations produce a data cube with $0\farcs1\times0\farcs1$ spatial elements.  To maximize signal we used the IFU's PRISM/CLEAR mode, with a nominal resolving power of $30-330$ between 0.6~$\mu$m -- 5.3~$\mu$m \citep{boker2023_nirspec}. Our four exposures made use of the 4-POINT-DITHER pattern and were read out with NRSIRS2RAPID, leading to a total exposure time of 2976~s.

Finally, supporting ground-based optical observations were obtained with
the Inamori Magellan Areal Camera and Spectrograph \citep[IMACS;][]{dressler2011_imacs} on the 6.5~m Magellan-Baade telescope at Las Campanas in Chile;
the Gemini Multi-Object Spectrograph - North \citep[GMOS-N;][]{hook2004_gmos} in imaging mode on the 8.1~m Gemini North (Gemini-N) telescope (program GN-2024B-Q-114) on Maunakea in Hawaii, USA;
the Gemini Multi-Object Spectrograph - South \citep[GMOS-S;][]{gimeno2016_gmoss} in imaging mode on the 8.1~m Gemini South (Gemini-S) telescope (programs GS-2023A-LP-104, GS-2024A-Q-111, and GS-2024B-Q-113) at Cerro Pach{\'o}n in Chile (program 2023A-396684);
the Large Monolithic Imager \citep[LMI;][]{bida2014_dct} on Lowell Observatory's 4.3~m Lowell Discovery Telescope (LDT; formerly named the Discovery Channel Telescope) at Happy Jack, Arizona, USA;
and
the European Southern Observatory (ESO) Faint Object Spectrograph and Camera \citep[EFOSC2;][]{buzzoni1984_efosc} on ESO's
3.58~m New Technology Telescope (NTT; program 113.26J9.002) at La Silla in Chile.
All ground-based observations reported here were obtained using Sloan $r'$-band filters.
Details of all instrumentation are shown in Table~\ref{table:instrumentation}, while observational circumstances of all ground-based observations are listed in Table~\ref{table:ground_observations_133p}.

\setlength{\tabcolsep}{6.5pt}
\setlength{\extrarowheight}{0em}
\begin{table*}[htb]
\caption{Ground-based $r'$-band Observations of 133P$^a$}
\centering
\smallskip
\footnotesize
\begin{tabular}{lcrrrrrrrrrr}
\hline\hline
\multicolumn{1}{c}{UT Date}
 & \multicolumn{1}{c}{Telescope$^b$}
 & \multicolumn{1}{c}{$N$$^c$}
 & \multicolumn{1}{c}{$t$$^d$}
 & \multicolumn{1}{c}{$\theta_s$$^e$}
 & \multicolumn{1}{c}{$\nu$$^f$}
 & \multicolumn{1}{c}{$r_h$$^g$}
 & \multicolumn{1}{c}{$\Delta$$^h$}
 & \multicolumn{1}{c}{$\alpha$$^i$}
 & \multicolumn{1}{c}{PA$_{-\odot}$$^j$}
 & \multicolumn{1}{c}{PA$_{-v}$$^k$}
 & \multicolumn{1}{c}{$\Delta t_q$$^l$}
 \\
\hline
2023 Apr 20 & Gemini-S &  2 &   600 & 0.8 & 274.9 & 3.048 & 2.045 &  1.4 & 131.4 & 290.2 & $-$386 \\ % 2460054.8
2023 Apr 22 & Palomar  &  42 & 12600 & 1.3 & 275.2 & 3.045 & 2.044 &  2.2 & 124.4 & 290.3 & $-$384 \\ % 2460056.8
2023 Jul 16 & Gemini-S &   3 &   900 & 0.9 & 291.7 & 2.919 & 2.729 & 20.4 & 111.6 & 290.4 & $-$299 \\ % 2460141.8
2024 Mar 07 & Magellan &  3 &   900 & 1.0 & 344.5 & 2.684 & 3.158 & 17.2 & 260.9 & 261.1 & $-$64 \\ % 2460376.9
2024 Mar 08 & Magellan &   1 &   230 & 2.2 & 344.8 & 2.683 & 3.147 & 17.4 & 260.8 & 261.0 & $-$63 \\ % 2460377.9
2024 Mar 16 & Gemini-S &   4 &   600 & 1.0 & 346.7 & 2.680 & 3.058 & 18.5 & 259.5 & 259.9 & $-$55  \\ % 2460385.9
2024 Apr 06 & Gemini-S &  2 &   300 & 1.1 & 351.8 & 2.674 & 2.807 & 20.9 & 256.6 & 257.5 & $-$34  \\ % 2460406.8
2024 Apr 07 & Gemini-S &   6 &   900 & 0.7 & 352.0 & 2.674 & 2.795 & 20.9 & 256.5 & 257.4 &  $-$33 \\ % 2460407.8
2024 May 08 & LDT &    6 &  1800 & 1.9 & 359.5 & 2.671 & 2.399 & 22.1 & 253.3 & 254.9 &  $-$2 \\ % 2460438.8
2024 May 10 & \textit{Perihelion} & --- & --- & --- & 0.0 & 2.671 & 2.368 & 22.3 & 253.1 & 254.7 & 0 \\  % 2460440.8
2024 May 11 & LDT &  3 &   900 & 1.9 &   0.3 & 2.671 & 2.360 & 22.1 & 253.1 & 254.7 & $+$1 \\ % 2460441.8
2024 Jun 01 & LDT      &  4 &  1200 & 1.2 &   5.4 & 2.672 & 2.104 & 20.4 & 251.8 & 253.8 & $+$22 \\ % 2460462.8
2024 Jun 07 & Gemini-S &  5 &   725 & 0.8 &   6.8 & 2.673 & 2.036 & 19.5 & 251.5 & 253.7 & $+$28 \\ % 2460468.8
2024 Jul 01 & Gemini-S &  7 &   630 & 0.8 &  12.6 & 2.679 & 1.811 & 13.8 & 250.7 & 253.6 & $+$52 \\ % 2460492.8
2024 Jul 15 & Gemini-S &  7 &   630 & 0.9 &  15.9 & 2.685 & 1.724 &  8.9 & 250.0 & 253.9 & $+$66 \\ % 2460506.8
2024 Aug 02 & Gemini-N &  8 &   720 & 0.7 &  20.3 & 2.693 & 1.680 &  1.3 & 236.1 & 254.7 & $+$84 \\ % 2460524.8
2024 Aug 09 & NTT      &  6 &  1240 & 2.5 &  21.9 & 2.697 & 1.685 &  1.7 &  86.5 & 255.1 & $+$91  \\ % 2460531.8
2024 Aug 10 & NTT      & 17 &  1150 & 0.9 &  22.1 & 2.698 & 1.687 &  2.1 &  83.7 & 255.1 & $+$92 \\ % 2460532.8
2024 Aug 11 & NTT      & 34 &  8684 & 1.3 &  22.4 & 2.698 & 1.690 &  2.6 &  81.7 & 255.2 & $+$93 \\ % 2460533.8
2024 Aug 27 & Gemini-N &  8 &   720 & 0.7 &  26.2 & 2.708 & 1.760 &  9.1 &  76.5 & 255.9 & $+$109 \\ % 2460549.8
2024 Sep 30 & Gemini-N &  9 &   810 & 0.7 &  34.2 & 2.734 & 2.086 & 18.4 &  75.4 & 256.4 & $+$143 \\ % 2460583.8
2024 Oct 03 & Magellan &  5 &   750 & 0.9 &  34.8 & 2.737 & 2.121 & 18.8 &  75.3 & 256.4 & $+$146 \\ % 2460586.8
2024 Oct 06 & Gemini-N &  9 &   810 & 0.7 &  35.6 & 2.740 & 2.162 & 19.3 &  75.2 & 256.3 & $+$149 \\ % 2460589.8
2024 Oct 26 & LDT      &  3 &   900 & 1.2 &  40.1 & 2.758 & 2.432 & 20.9 &  74.1 & 255.4 & $+$169 \\ % 2460609.8
2024 Oct 27 & Gemini-S &  9 &   810 & 0.7 &  40.3 & 2.759 & 2.445 & 20.9 &  74.1 & 255.4 & $+$170 \\ % 2460610.8
2024 Oct 28 & Gemini-S &  8 &   720 & 0.7 &  40.5 & 2.760 & 2.459 & 20.9 &  74.0 & 255.3 & $+$171 \\ % 2460611.8
2024 Nov 05 & LDT      &  3 &   900 & 1.2 &  42.4 & 2.768 & 2.575 & 21.0 &  73.4 & 254.8 & $+$179 \\ % 2460619.8
2024 Nov 22 & Gemini-N & 10 &   900 & 0.6 &  46.2 & 2.786 & 2.821 & 20.3 &  72.2 & 253.6 & $+$196 \\ % 2460636.8
\hline
\hline
\multicolumn{12}{l}{$^a$ Observing geometry parameters from JPL Horizons (Orbit solution \#77) \citep{giorgini1996_horizons}} \\
\multicolumn{12}{l}{$^b$ See Table~\ref{table:instrumentation} for explanations of telescope designations.} \\
\multicolumn{12}{l}{$^c$ Number of usable exposures.} \\
\multicolumn{12}{l}{$^d$ Total exposure time of usable exposures, in s.} \\
\multicolumn{12}{l}{$^e$ FWHM seeing, in arcseconds.} \\
\multicolumn{12}{l}{$^f$ True anomaly, in degrees.} \\
\multicolumn{12}{l}{$^g$ Heliocentric distance, in au.} \\
\multicolumn{12}{l}{$^h$ Geocentric distance, in au.} \\
\multicolumn{12}{l}{$^i$ Solar phase angle (observer-target-Sun), in degrees.} \\
\multicolumn{12}{l}{$^j$ Position angle of the anti-Solar vector as projected on the sky, in degrees East of North.} \\
\multicolumn{12}{l}{$^k$ Position angle of the negative heliocentric velocity vector as projected on the sky, in degrees East of North.} \\
\multicolumn{12}{l}{$^l$ Time prior to (negative values) or after (positive values) perihelion, in days.} \\
\end{tabular}
\label{table:ground_observations_133p}
\end{table*}

All ground-based observations were conducted using non-sidereal tracking and at airmasses of $\lesssim2.5$, with typical seeing conditions of $\sim1''-2''$.  A minimum of three exposures was obtained during each visit in order to ensure that our target and any associated activity could be unambiguously identified from their non-sidereal motion.  On three of our 27 nights of observing (UT 2023 April 20, UT 2024 March 8, and UT 2024 April 6), however, one or more detections were discarded due to being too close to background sources for photometry to be considered reliable (see Sections~\ref{section:optical_data_processing} and \ref{section:optical_photometry}), leading to fewer than three exposures being reported for those nights (see Table~\ref{table:ground_observations_133p}).

\section{Data Processing\label{section:data_processing}}

\subsection{NIRSpec Data\label{section:nirspec_processing}}

\begin{figure*}[htb]
    \centering
    \includegraphics[width=0.47\linewidth]{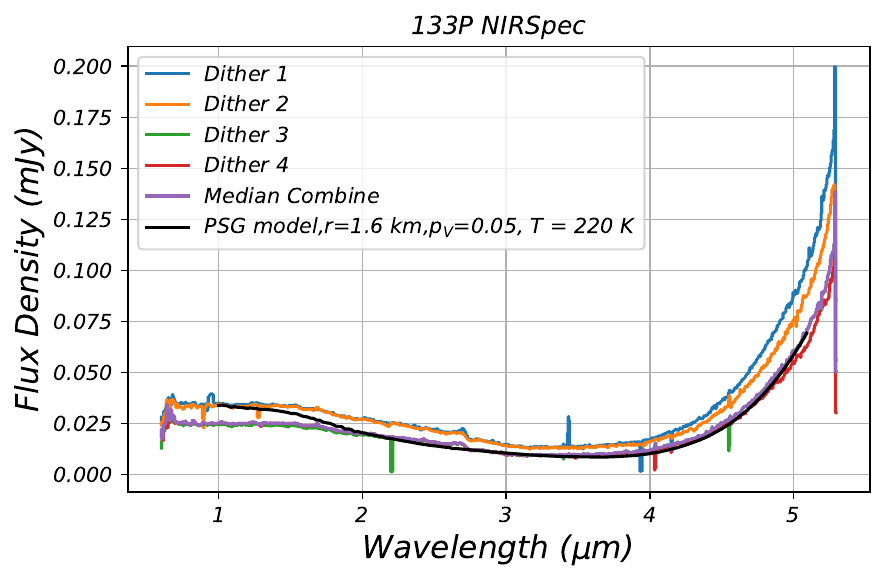}
    \includegraphics[width=0.47\linewidth]{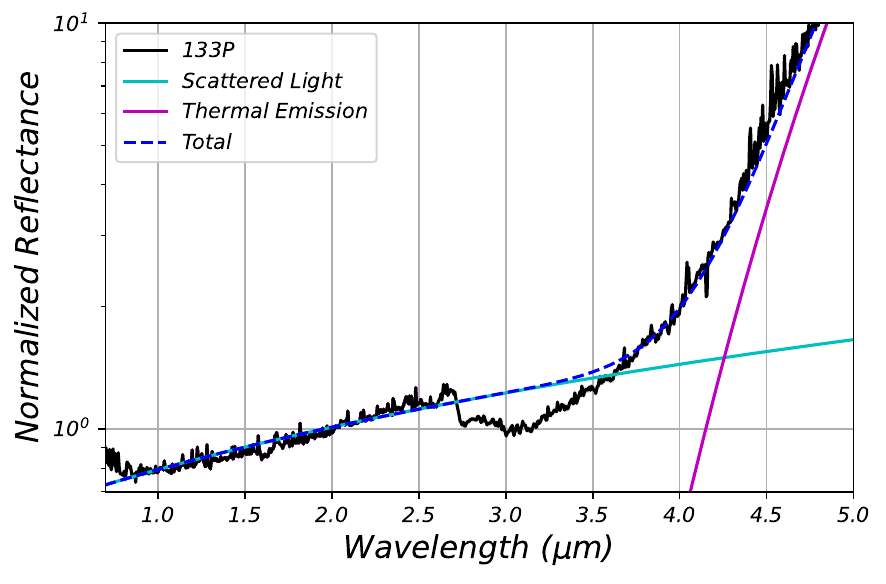}
    \caption{Left: NIRSpec observations of 133P/Elst-Pizarro obtained on UT 2024 June 12 (\jwst{} Cycle 2) for each of four NIRSpec dithers, compared to a bare-nucleus model ($p_V=0.05$, $T = 220$~K, $r=1.6$~km), generated with  the Planetary Spectrum Generator for the corresponding observing geometry. Individual dithers were median-combined to produce a higher signal-to-noise spectrum for analysis. Right: Normalized reflectance after dividing by the solar spectrum. Shown are a linear reflectance model for scattered light (cyan) and a 180~K black body emission curve normalized at 4.14~$\mu$m (magenta), with the combined contribution represented by a blue dotted line.  }
    \label{fig:Cyc2_NIRSpec_raw_ref}
\end{figure*}

\begin{figure*}[htb]
    \centering
    \includegraphics[width=0.47\linewidth]{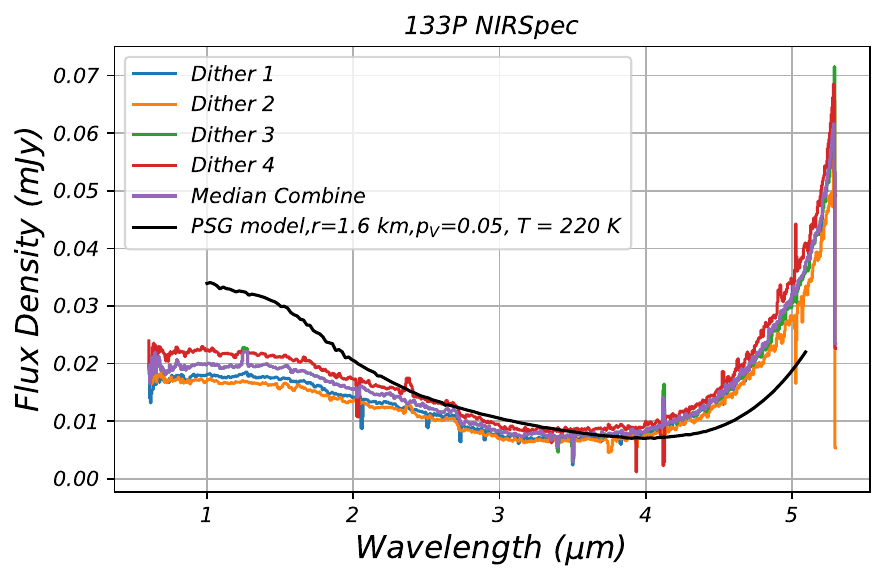}
    \includegraphics[width=0.47\linewidth]{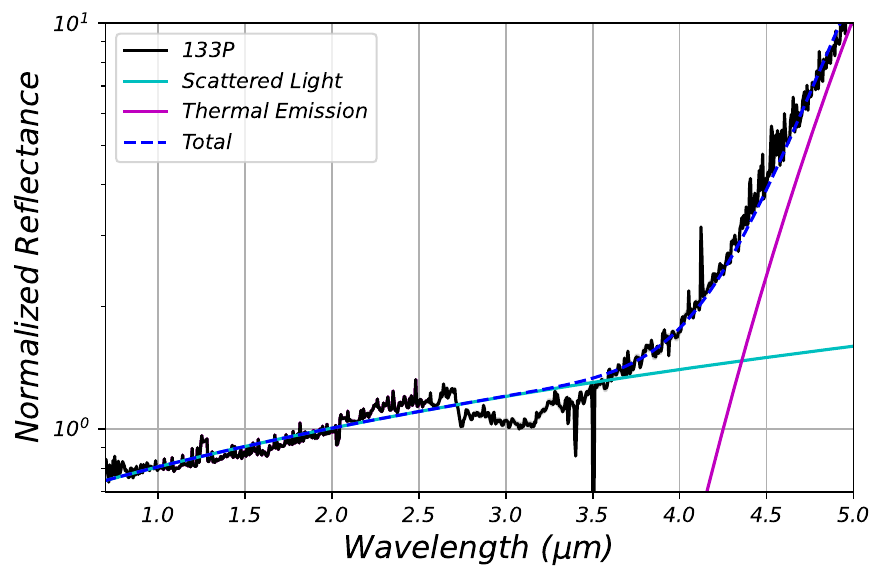}
    \caption{NIRSpec observations of 133P/Elst-Pizarro obtained on UT 2024 September 20 (\jwst{} Cycle 3), with dither and median combined spectra on the left and normalized reflectance on the right. See the caption of Figure~\ref{fig:Cyc2_NIRSpec_raw_ref} for more detailed plot descriptions. }
    \label{fig:Cyc3_NIRSpec_raw_ref}
\end{figure*}

To calibrate our NIRSpec observations of 133P, we also acquired off-source background sky frames taken 1\arcmin{} away, which were used for subtraction with the \jwst{} {\tt python} package and the {\tt spec2} data reduction pipeline\footnote{\url{https://jwst-pipeline.readthedocs.io/}} \citep{bushouse2023_jwstcalibrationpipeline_1_12_5}. The resulting background spectrum was removed from the IFU data; because the comet's extended coma fills much of the field, we could not ensure that a dust- and/or gas-free background region existed within the $3\arcsec\times3\arcsec$ FOV of the on-source observations. The {\tt spec2} pipeline was executed using version 1.12.5, which allowed application of the NSClean step to mitigate the $1/f$ noise feature \citep{rauscher2024_nsclean}. At this point, the pipeline photometrically calibrated the data to radiance units of MJy~sr$^{-1}$.

A one-dimensional spectrum was extracted from each dither using a circular aperture with a radius of $0\farcs4$ (Figures~\ref{fig:Cyc2_NIRSpec_raw_ref} and \ref{fig:Cyc3_NIRSpec_raw_ref}). This extraction employed a slice-by-slice procedure in {\tt python}, centering the aperture on the brightest pixel near the comet's predicted position in each frame. By doing so, we minimized the impact of small, wavelength-dependent shifts in the IFU optocenter at the pixel scale. To prevent extended ``snowball'' cosmic ray events \citep{regan2024_snowballs} --- which were sometimes not removed by earlier cosmic ray cleaning steps --- from biasing the optocenter determination, we applied an additional routine that identified all pixels lying $50\sigma$ above the median value of their immediate neighbors. The fluxes in these pixels were then replaced with the median of their nearest neighboring pixels. This approach successfully flagged extended cosmic ray features while preserving genuine dust coma emission. After applying these two custom procedures, the resulting one-dimensional spectra from the dataset were found to be nearly identical in absolute flux, making them well suited for stacking to search for faint gas emission lines. The full process was carried out independently for both 133P NIRSpec visits listed in Table \ref{table:jwst_mbc_observations}.

\begin{figure*}[ht]
    \centering
    \includegraphics[width=0.49\linewidth]{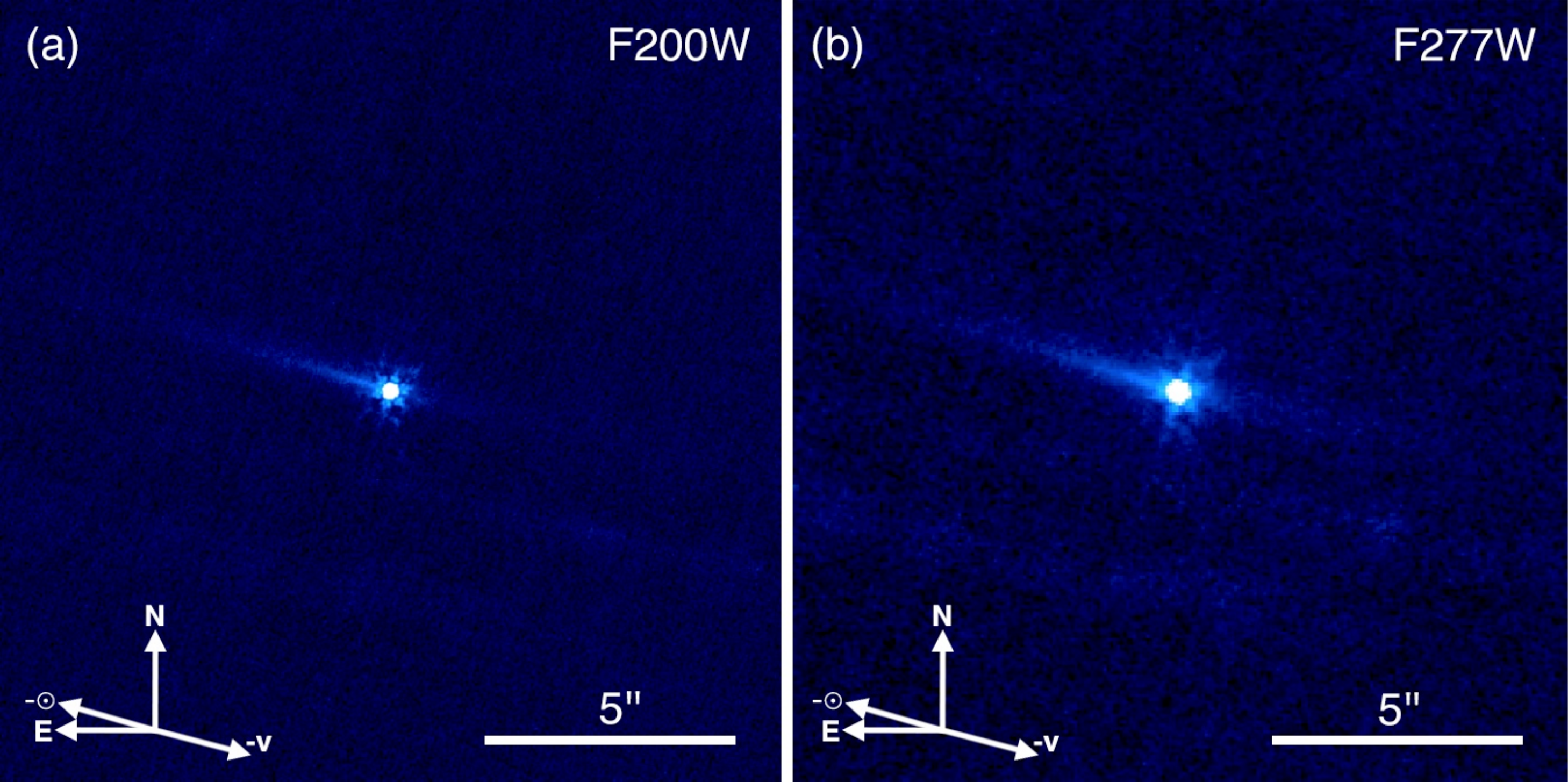}~~~
    \includegraphics[width=0.49\linewidth]{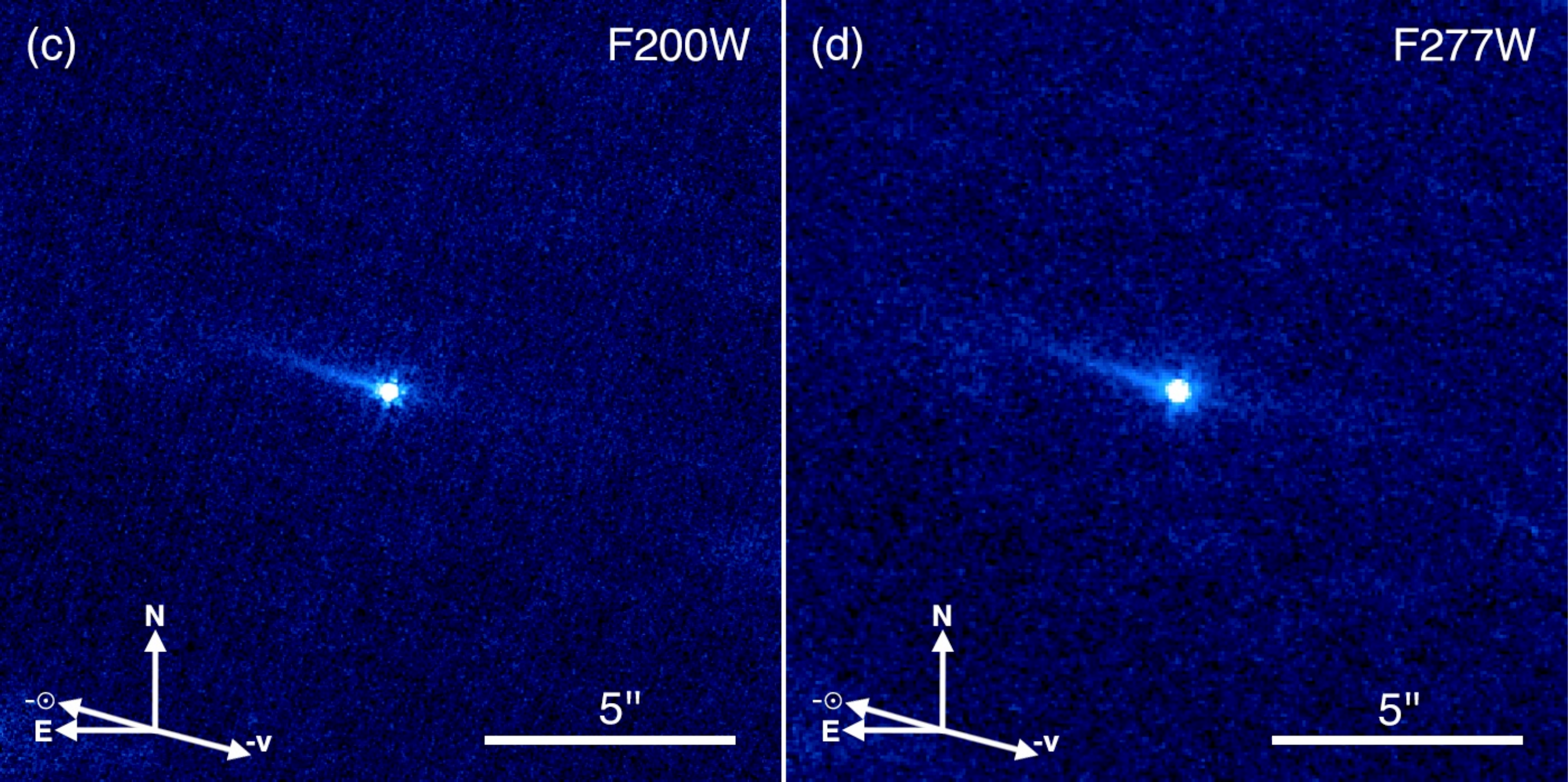}
    \caption{Median composite images of 133P/Elst-Pizarro, aligned on the photocenter of the comet in each individual image, constructed from (a) F200W and (b) F277W NIRCam data obtained on UT 2024 October 14, and (c) F200W and (d) F277W NIRCam data obtained on UT 2024 October 28, comprising 1031~s of total exposure time each.  Labeled arrows indicate the directions of celestial north (N) and east (E), and the projected anti-Sun ($-\odot$) and negative heliocentric velocity ($-v$) vectors as seen from \jwst{}. A $5''$ angular scale bar (8200 km and 9000~km at the distance of the comet on UT 2024 October 14 and UT 2024 October 28, respectively) is also shown in each panel. Color scaling in both panels is logarithmic, where regions in the inner coma (in the center of each image) that are shown as solid white consist of pixels with fluxes that are $\sim25$\% of the peak central flux or larger.}
    \label{fig:nircam_images}
\end{figure*}

\subsection{NIRCam Data\label{section:nircam_data_processing}}

NIRCam images were processed with pipeline version 12.0.5 and Calibration Reference Data System context file number 1303.  
At the time of these observations, individual pipeline-processed NIRCam images had a visible horizontal ``striped'' background structure that is attributed to $1/f$ noise from \jwst{}'s SIDECAR ASICs detector readout electronics\footnote{\url{https://jwst-docs.stsci.edu/known-issues-with-jwst-data/nircam-known-issues/nircam-1-f-noise-removal-methods}}, as well as relatively large numbers of randomly distributed pixels with {\tt NaN} values.

To reduce the impact of these noise features on subsequent analyses, we followed the process described in \citet{hsieh2025_358p}, where we first used the {\tt image1overf} package developed by C.\ Willott\footnote{\url{https://github.com/chriswillott/jwst}} to remove row median levels while preserving overall flux levels in order to reduce the background structure caused by $1/f$ noise.
We then used our own {\tt python} code to replace {\tt NaN} pixel values (which comprised $\sim1$\% of all pixels in the uncorrected images) and uncertainties with the median value and uncertainty of all adjacent non-{\tt NaN} pixels.

To enable better characterization of the detailed morphology of the comet and to mitigate the effect of cosmic rays and detector artifacts, we used {\tt pyraf}\footnote{\url{https://pypi.org/project/pyraf/}} \citep{stsci2012_pyraf} to construct median composite images of the object in each filter. This was done by shifting and aligning individual images on the object's photocenter using linear interpolation and performing a median combination of the resulting images for each filter.  Final median composite images are shown in Figure~\ref{fig:nircam_images}.

\subsection{Ground-Based Optical Data\label{section:optical_data_processing}}

Standard bias subtraction, flatfield correction, and cosmic ray removal were performed for all optical images obtained from ground-based facilities.  This reduction was carried out using {\tt python} code utilizing the {\tt ccdproc} package\footnote{\url{https://ccdproc.readthedocs.io/}} \citep{craig2023_ccdproc} in \citep{astropy2018_astropy} and the {\tt L.A.Cosmic} {\tt python} module\footnote{\url{https://pypi.org/project/lacosmic/}} \citep{vandokkum2001_lacosmic,vandokkum2012_lacosmic}.

\setlength{\tabcolsep}{6.0pt}
\setlength{\extrarowheight}{0em}
\begin{table*}[htb]
\caption{Photometric Results for 133P}
\centering
\smallskip
\footnotesize
\begin{tabular}{lccccccccc}
\hline\hline
\multicolumn{1}{c}{UT Date}
 & \multicolumn{1}{c}{Telescope$^a$}
 & \multicolumn{1}{c}{$\rho_{0}$$^b$}
 & \multicolumn{1}{c}{$d_{0}$$^c$}
 & \multicolumn{1}{c}{$m_{r'}(r_h,\Delta,\alpha)$$^d$}
 & \multicolumn{1}{c}{$m_{V}(1,1,0)$$^e$}
 & \multicolumn{1}{c}{$A_d/A_n$$^f$}
 & \multicolumn{1}{c}{$Af\rho_{0}$$^g$}
 & \multicolumn{1}{c}{$\rho_{5000{\rm km}}$$^h$}
 & \multicolumn{1}{c}{$Af\rho_{5000{\rm km}}$$^i$}
 \\
\hline
2023 Apr 20 & Gemini-S & 1.9 & 2800 & 19.88$\pm$0.01 & 15.87$\pm$0.03 & 0.01$\pm$0.04 & 0.3$\pm$1.0 & 3.4 & 0.2$\pm$1.0 \\
2023 Apr 22 & Palomar  & 3.0 & 4400 & 20.1$\pm$0.2 & 16.0$\pm$0.2 & 0.0 & 0.0 & 3.4 & 0.0 \\
2023 Jul 16 & Gemini-S & 1.9 & 3800 & 21.20$\pm$0.02 & 15.80$\pm$0.05 & 0.08$\pm$0.06 & 1.7$\pm$1.2 & 2.5 & 1.3$\pm$1.4 \\
2024 Mar 07 & Magellan & 2.0 & 4600 & 21.1$\pm$0.6 & 15.7$\pm$0.4 & 0.2$\pm$0.5 & 3.3$\pm$8.8 & 2.2 & 3.1$\pm$8.9 \\
2024 Mar 08 & Magellan & 2.0 & 4600 & 21.25$\pm$0.01 & 15.82$\pm$0.05 & 0.05$\pm$0.05 & 1.0$\pm$0.9 & 2.2 & 0.9$\pm$1.0 \\
2024 Mar 16 & Gemini-S & 2.4 & 5300 & 21.25$\pm$0.07 & 15.84$\pm$0.07 & 0.04$\pm$0.07 & 0.6$\pm$1.1 & 2.3 & 0.6$\pm$1.1 \\
2024 Apr 06 & Gemini-S & 2.6 & 4300 & 21.02$\pm$0.01 & 15.75$\pm$0.05 & 0.13$\pm$0.06 & 2.0$\pm$0.9 & 2.5 & 2.2$\pm$1.4 \\
2024 Apr 07 & Gemini-S & 1.6 & 3200 & 21.08$\pm$0.02 & 15.80$\pm$0.05 & 0.08$\pm$0.06 & 2.0$\pm$1.4 & 2.5 & 1.5$\pm$1.6 \\
2024 May 08$^*$ & LDT & 3.6 & 6300 & 20.9$\pm$0.1 & 15.9$\pm$0.1 & 0.0 & 0.0 & 2.9 & 0.0 \\
2024 May 11$^*$ & LDT & 3.1 & 5300 & 20.60$\pm$0.02 & 15.69$\pm$0.05 & 0.19$\pm$0.06 & 3.0$\pm$0.9 & 2.9 & 3.3$\pm$1.9 \\
2024 Jun 01$^*$ & LDT      & 2.6 & 4000 & 20.6$\pm$0.1 & 15.9$\pm$0.1 & 0.0 & 0.0 & 3.3 & 0.0 \\
2024 Jun 07$^*$ & Gemini-S & 2.1 & 3100 & 20.31$\pm$0.01 & 15.78$\pm$0.05 & 0.10$\pm$0.06 & 2.7$\pm$1.4 & 3.4 & 2.1$\pm$1.8 \\
2024 Jul 01$^*$ & Gemini-S & 2.4 & 3200 & 20.04$\pm$0.05 & 15.93$\pm$0.08 & 0.0 & 0.0 & 3.8 & 0.0 \\
2024 Jul 15$^*$ & Gemini-S & 2.4 & 3000 & 19.78$\pm$0.04 & 15.97$\pm$0.06 & 0.0 & 0.0 & 4.0 & 0.0 \\
2024 Aug 02 & Gemini-N & 1.9 & 2300 & 19.31$\pm$0.01 & 16.01$\pm$0.02 & 0.0 & 0.0 & 4.1 & 0.0 \\
2024 Aug 09 & NTT      & 5.5 & 6700 & 18.96$\pm$0.07 & 15.64$\pm$0.07 & 0.24$\pm$0.08 & 3.0$\pm$1.0 & 4.1 & 3.9$\pm$2.2 \\
2024 Aug 10 & NTT      & 2.2 & 2700 & 19.20$\pm$0.08 & 15.79$\pm$0.08 & 0.09$\pm$0.08 & 2.7$\pm$2.3 & 4.1 & 1.7$\pm$2.5 \\
2024 Aug 11 & NTT      & 6.0 & 4400 & 19.13$\pm$0.08 & 15.73$\pm$0.07 & 0.15$\pm$0.08 & 1.6$\pm$0.9 & 4.1 & 2.5$\pm$1.6 \\
2024 Aug 27 & Gemini-N & 2.4 & 3100 & 19.63$\pm$0.07 & 15.78$\pm$0.06 & 0.10$\pm$0.07 & 2.7$\pm$1.9 & 3.9 & 1.6$\pm$2.1 \\
2024 Sep 30$^{**}$ & Gemini-N & 2.4 & 3600 & 20.22$\pm$0.02 & 15.66$\pm$0.04 & 0.23$\pm$0.06 & 5.1$\pm$1.2 & 3.3 & 3.7$\pm$2.3 \\
2024 Oct 03$^{**}$ & Magellan & 2.4 & 3700 & 20.46$\pm$0.03 & 15.80$\pm$0.05 & 0.07$\pm$0.06 & 1.6$\pm$1.3 & 3.3 & 1.2$\pm$1.4 \\
2024 Oct 06$^{**}$ & Gemini-N & 1.6 & 2500 & 20.46$\pm$0.02 & 15.75$\pm$0.07 & 0.12$\pm$0.06 & 4.1$\pm$1.8 & 3.2 & 2.0$\pm$2.1 \\
2024 Oct 26$^{**}$ & LDT      & 2.4 & 4200 & 20.64$\pm$0.07 & 15.65$\pm$0.08 & 0.24$\pm$0.06 & 4.2$\pm$1.1 & 2.8 & 3.5$\pm$2.1 \\
2024 Oct 27$^{**}$ & Gemini-S & 1.9 & 3400 & 20.65$\pm$0.01 & 15.65$\pm$0.04 & 0.24$\pm$0.06 & 5.8$\pm$1.3 & 2.8 & 4.1$\pm$2.5 \\
2024 Oct 28$^{**}$ & Gemini-S & 1.9 & 3400 & 20.88$\pm$0.07 & 15.81$\pm$0.07 & 0.07$\pm$0.08 & 1.7$\pm$1.8 & 2.8 & 1.2$\pm$1.9 \\
2024 Nov 05 & LDT      & 2.6 & 4900 & 20.90$\pm$0.01 & 15.75$\pm$0.07 & 0.13$\pm$0.06 & 2.3$\pm$0.9 & 2.7 & 2.1$\pm$1.4 \\
2024 Nov 22 & Gemini-N & 1.9 & 3900 & 21.12$\pm$0.08 & 15.77$\pm$0.09 & 0.11$\pm$0.08 & 2.3$\pm$1.7 & 2.4 & 1.8$\pm$2.0 \\
\hline
\hline
\multicolumn{10}{l}{$^a$ See Table~\ref{table:instrumentation} for explanations of telescope designations.} \\
\multicolumn{10}{l}{$^b$ Aperture radius, in arcseconds, determined to be optimal for photometry measurements from a curve of growth analysis.} \\
\multicolumn{10}{l}{$~~~~$growth analysis.} \\
\multicolumn{10}{l}{$^c$ Distance, in km, equivalent to $\rho_0$ at the geocentric distance of the comet.} \\
\multicolumn{10}{l}{$^d$ Mean $r'$-band apparent magnitude using photometry apertures with $\rho_0$.} \\
\multicolumn{10}{l}{$^e$ Mean $V$-band absolute magnitude (normalized to $r_h=\Delta=1$~au and $\alpha=0^{\circ}$) using photometry apertures with $\rho_0$.} \\
\multicolumn{10}{l}{$^f$ Inferred dust-to-nucleus scattering surface area ratio} \\
\multicolumn{10}{l}{$^g$ $Af\rho$, in cm, computed from photometry measured using apertures with $\rho_0$.} \\
\multicolumn{10}{l}{$^h$ Angular equivalent, in arcseconds, to a $\rho=5000$~km photometry aperture at the geocentric distance of the comet.} \\
\multicolumn{10}{l}{$^i$ $Af\rho$, in cm, for $\rho=5000$~km computed from a best-fit power law to measured $Af\rho$ vs.\ aperture data, with } \\
\multicolumn{10}{l}{$~~~~$uncertainties incorporating the range of $Af\rho$ values for $\rho=(5000\pm2000)$~km.} \\
\multicolumn{10}{l}{$^*$ Photometry used to determine average $Af\rho$ at the time of the UT 2024 June 12 NIRSpec visit.} \\
\multicolumn{10}{l}{$^{**}$ Photometry used to determine average $Af\rho$ at the time of the UT 2024 October 14 NIRSpec visit.} \\
\end{tabular}
\label{table:ground_photometry_133p}
\end{table*}

To maximize signal-to-noise ratios for morphological analyses (see Section~\ref{section:morphology_optical}) and background source contamination assessment for photometric measurements (see Section~\ref{section:optical_photometry}), we constructed composite images of the object for each night of data by shifting and aligning individual images on the object's photocenter using linear interpolation and then co-adding the images together (Figure~\ref{fig:133p_optical_images}).  Composite images were also constructed by aligning individual images from each night on samples of moderately bright field stars and adding the images together in order to enhance the visibility of faint stationary background sources near detections of the comet that could potentially contaminate photometric measurements of the comet.

\section{Results and Analysis}\label{section:results_analysis}

\subsection{Spectroscopy}\label{section:spectroscopy}

\subsubsection{H$_2$O Production}\label{section:H2O_analysis}
To derive water production rates from our two epochs, we first obtain model spectra of the H$_2$O emission spectrum across our NIRSpec bandpass for the appropriate observing conditions in Table \ref{table:jwst_mbc_observations} for a production rate of $5\times10^{24}$ molecules~s$^{-1}$. Using the NASA Planetary Spectrum Generator\footnote{\url{https://psg.gsfc.nasa.gov/}} \citep[PSG;][]{villanueva2018_psg} and its API functionality, we retrieved model spectra for the nucleus, dust, and gases of interest (H$_2$O, CO$_2$, CO, and CH$_3$OH) in a 100:1:1:10 mixture, using a best-fit effective gas temperature of $T_{eff}=35$~K as determined using the PSG's retrieval suite. We tested temperatures from 15 K to 60 K, with no significant changes to the derived H$_2$O production rate within uncertainties; our lower signal-to-noise across the H$_2$O feature makes it more difficult to constrain the temperature. This temperature is slightly higher than, but consistent with, the two-temperature fit of 15 K and 30 K used for 358P \citep{hsieh2025_358p}. The PSG takes inputs for heliocentric distance, heliocentric velocity, production rate, and coma gas temperature to generate a simulated ro-vibrational spectrum for molecular species using the Cometary Emission Model to produce emission efficiencies for molecular emission features \citep{Villanueva2011,villanueva2018_psg}, and calculates the observed flux for the provided user geometry.

For the complete description of the equations and methodology, see Equations 32-44 in \citet{villanueva2018_psg}. The PSG uses the Haser model \citep{haser1957_cometmodel} to determine the column density of H$_2$O as a function of impact parameter, assuming photo-destruction lifetimes from \citet{huebner_photoionization_2015}, and from this calculate the number of molecules within the instrument beam. By combining the number of molecules in the beam with the line-by-line emission efficiencies for the provided observing conditions and initial production rates, the PSG is able to produce a model emission spectrum. The modeled spectrum in wavelength space can be defined as, 
\begin{equation}
    F_\lambda(Q,r_\Delta,r_h,\dot{r_h},T) = \frac{N(Q,r_\Delta)g_\lambda(r_h,\dot{r_h},T)}{4\pi r_\Delta^2}
\end{equation}
where $N$ is the total number of molecules in the aperture as defined by the Haser model and $g_\lambda$ is the excitation rate for the individual ro-vibrational lines of the transition in photons molecules$^{-1}$ sec$^{-1}$. 
The PSG is also capable of retrieving fits of molecular production rate and temperature to the data as well.

Following \citet{hsieh2025_358p}, we export the synthetic PSG H$_2$O spectrum to perform more customized fitting routines in Python.  Using the {\tt scipy.optimize.curvefit} function, we perform an initial polynomial best fit to the 2.5 - 3.0 $\mu$m region, and supply this to the {\tt emcee} package to explore correlated error in the fit continuum and derived H$_2$O production rates, similar to \citet{hsieh2025_358p}. We initialize 100 walkers and allow them to run for 15000 iterations, using a burn-in of 1500 iterations to derive the uncertainty on our model fits.

\begin{figure*}
\centering
\includegraphics[width=0.47\linewidth]{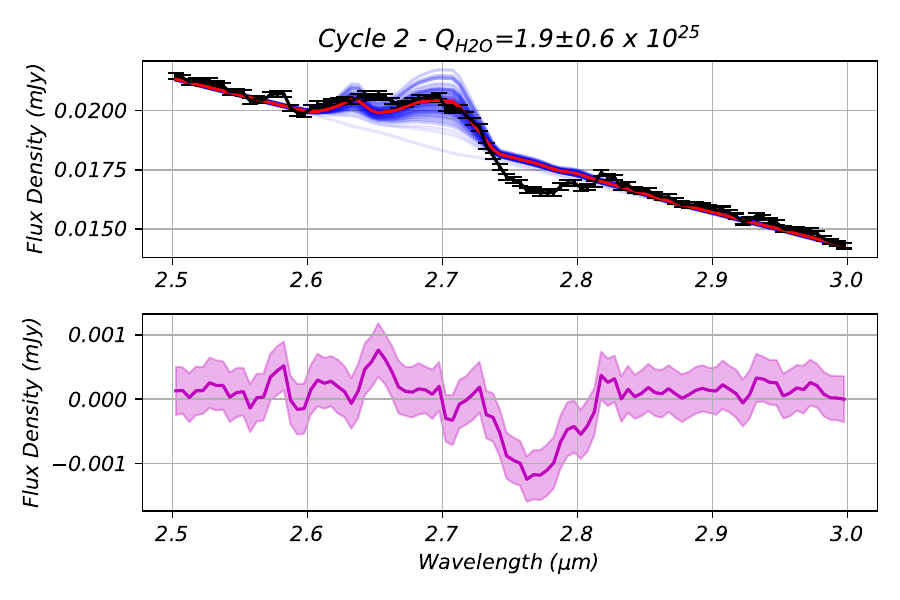}
\includegraphics[width=0.47\linewidth]{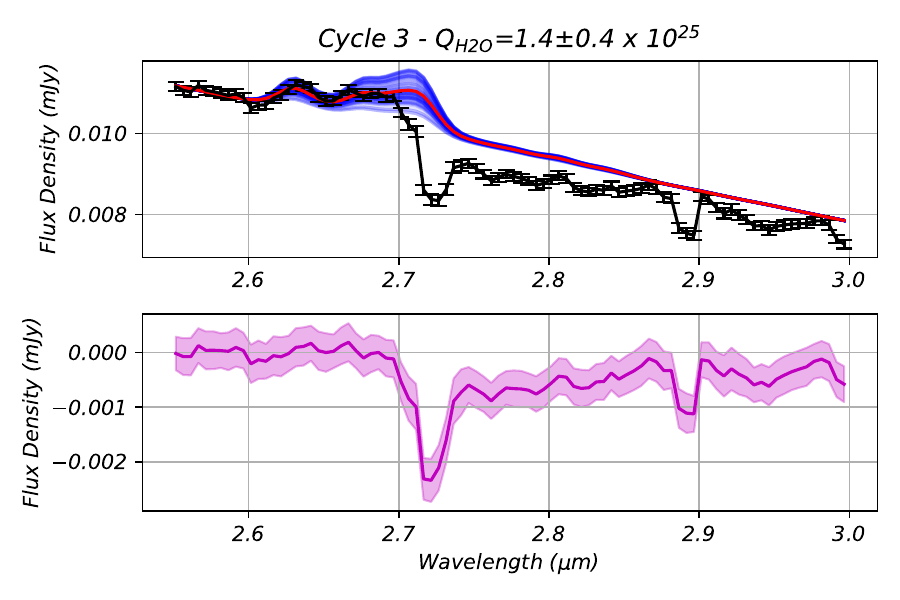}
\caption{PSG model fits (red) with residuals to the H$_2$O emission of 133P on UT 2024 June 12 (Cycle 2) and 2024 October 14 (Cycle 3). A continuum is fit simultaneously with a Markov Chain Monte Carlo model using the \textit{emcee} Python package to characterize uncertainties in the production rate fit, and a random subset of walkers are shown in blue. Our H$_2$O emission model fits use $T_{\rm eff}=35$~K, but we note that effects from temperature differences are captured within our derived uncertainties. We note an apparent significant absorption feature just redward of 2.72~$\mu$m that could depress our retrieved production rates; the best fit for Cycle 3 is when we omit the 2.70$-$2.72~$\mu$m region. The changing bandwidth between Cycle 2 and 3 makes it difficult to correct for this accordingly.}
\label{fig:133P_H2O_production}
\end{figure*}

We find $Q_{\rm H_2O}=(1.9\pm0.6)\times10^{25}$~molecules~s$^{-1}$, equivalent to $(0.6\pm0.2)$~kg~s$^{-1}$, on UT 2024 June 12 (in Cycle 2), and a slightly lower value of $Q_{\rm H_2O}=(1.4\pm0.4)\times10^{25}$~molecules~s$^{-1}$, equivalent to $(0.4\pm0.1)$~kg~s$^{-1}$, on UT 2024 September 20 (in Cycle 3). Fits and samples of walkers are shown in Figure \ref{fig:133P_H2O_production}.
These results correspond to a nominal 25\% decline in $Q_{\rm H_2O}$ between the two NIRSpec visits, which is within the range of predicted amounts of decline in the water sublimation rate ($\sim$10$-$25\%) due to the increase in heliocentric distance between the two visits (see below). Within uncertainties, however, we note that the measured $Q_{\rm H_2O}$ rates are also consistent with constant or even slightly increased water production between the two visits.  Importantly, however, water vapor was successfully detected during visits four months apart, confirming the action of a prolonged emission event.

As can be clearly seen in Figure \ref{fig:133P_H2O_production}, there is an absorption feature just redward of the H$_2$O emission in both datasets. The feature has different band centers (2.78 $\mu$m vs.\ 2.72 $\mu$m), band widths (0.1 $\mu$m vs.\ 0.18 $\mu$m) and in both cases has overlap with the H$_2$O emissions between 2.62~$\mu$m and 2.72~$\mu$m. This suggests that the the band shape of the H$_2$O emission feature beyond 2.7~$\mu$m may be dragged down by the absorption, and indeed our best production rate fits come from excluding the 2.70 $\mu$m -- 2.72 $\mu$m region. We point out that the higher H$_2$O production rate in the first set of observations may mask the sharp, narrow absorption feature at 2.71~$\mu$m, if due to surface material absorption. More details on possible sources for this narrow feature are discussed in Section \ref{section:refl_spectroscopy}.

Using the ice sublimation model developed by \citet{cowan1979_cometsublimation} that is available at the NASA Planetary Data System's Small Bodies Node\footnote{\url{https://pds-smallbodies.astro.umd.edu/tools/ma-evap/index.shtml}}, we can compute the expected water ice sublimation rate per unit area, ${\dot m_w}$, from the surface of a reference comet (rotating or non-rotating) at a given heliocentric distance. 
At $r_h=2.674$~au where 133P was first observed by NIRSpec, we find ${\dot m_w}=2.4\times10^{20}$~molecules~s$^{-1}$~m$^{-2}$ in the non-rotating or pole-on case (which produces the maximum attainable temperature for an object), and ${\dot m_w}=0.39\times10^{20}$~molecules~s$^{-1}$~m$^{-2}$ in the zero-obliquity, fast-rotating case.
Our derived best-fit water production rate at the time of our first NIRSpec observation (on UT 2024 June 12) then corresponds to an effective active area of $A_{\rm act}=(7.9\pm2.5)\times10^4$~m$^2$ and an active fraction of $f_{\rm act}=(1.6\pm0.9)\times10^{-3}$ (using $r_n=(2.0\pm0.5)$~km for 133P) in the non-rotating or pole-on case, and $A_{\rm act}=(4.9\pm1.5)\times10^5$~m$^2$ and $f_{\rm act}=(1.0\pm0.6)\times 10^{-2}$ in the zero-obliquity, fast-rotating case.

Meanwhile, at $r_h=2.747$~au where 133P was observed by NIRSpec for the second time, we find ${\dot m_{w}}\sim2.2\times10^{20}$~molecules~s$^{-1}$~m$^{-2}$ in the non-rotating or pole-on case and ${\dot m_{w}}\sim0.29\times10^{20}$~molecules~s$^{-1}$~m$^{-2}$ in the zero-obliquity, fast-rotating case.
Our derived best-fit water production rate of $Q_{\rm H_2O}=(1.4\pm0.4)\times10^{25}$~molecules~s$^{-1}$ at the time of our second NIRSpec observation (on UT 2024 October 14) then corresponds to effective active areas of $A_{\rm act}=(6.4\pm1.8)\times10^4$~m$^2$ and $f_{\rm act}=(1.3\pm0.7)\times10^{-3}$ in the non-rotating or pole-on case, and $A_{\rm act}=(4.8\pm1.4)\times10^5$~m$^2$ and $f_{\rm act}=(9.6\pm5.5)\times10^{-3}$ in the zero-obliquity, fast-rotating case.

\subsubsection{CO, CO$_2$ and CH$_3$OH Production Limits}\label{section:minor_species_analysis}

We do not see any evidence of the more volatile species CO, CO$_2$, or CH$_3$OH in our spectra, similar to the results reported for 238P and 358P \citep{kelley2023_jwst238p,hsieh2025_358p}. We use the same {\tt emcee} modeling technique described in \citet{hsieh2025_358p}, and above to establish the fitting uncertainties on emission features for each of the species, again using a spectrum queried from the PSG for relative abundances of 1\%, 1\%, and 10$\%$ at 35~K as initial guesses for CO, CO$_2$, and CH$_3$OH, respectively.

We constrain the 3$\sigma$ (99.7$\%$) upper limits on each of the volatiles by fitting polynomial continua and emission spectra of each species at the regions of the first excited ground state of CO (4.67~$\mu$m), the $\nu_3$ band of CO (4.26~$\mu$m), and the $\nu_2$ and $\nu_3$ bands of CH$_3$OH at 3.34~$\mu$m and 3.52~$\mu$m, and then allowing 1000 walkers to explore the parameter space over 15\,000 iterations. Our derived 3$\sigma$ upper limits for each epoch and volatile are reported in Table \ref{tab:133P_minor_volatiles}. We note that the CH$_3$OH upper limit is affected by a small absorption feature at 3.5~$\mu$m, that is not well captured with a polynomial continuum in both sets of observations. However, there is no sign of CH$_3$OH emission at 3.4~$\mu$m, and we therefore consider our upper limit to be accurate.  For reference, we also compute the corresponding upper limits to $Q_{\rm CO_2}/Q_{\rm H_2O}$ based on the water production rates computed in Section~\ref{section:H2O_analysis} and list these in the table as well.

\setlength{\tabcolsep}{3.5pt}
\setlength{\extrarowheight}{0em}
\begin{table}[htb]
\caption{Minor Volatile Species 3$\sigma$ Upper Limits for 133P$^a$}
\label{tab:133P_minor_volatiles}
\centering
\smallskip
\begin{tabular}{ccccc}
\hline\hline
\multicolumn{1}{c}{Cycle}
& \multicolumn{1}{c}{$Q_{\rm CO}$}
& \multicolumn{1}{c}{$Q_{\rm CO_2}$}
& \multicolumn{1}{c}{$Q_{\rm CH_3OH}$}
& \multicolumn{1}{c}{$Q_{\rm CO_2}/Q_{\rm H_2O}$}
\\
\hline
2 & $<$9.8$\times$10$^{24}$ & $<$1.4$\times$10$^{23}$ & $<$3.3$\times$10$^{23}$ & $<$0.007 \\
3 &  $<$5.5$\times$10$^{24}$ & $<$1.3$\times$10$^{23}$ & $<$4.9$\times$10$^{23}$ & $<$0.009 \\ 
\hline\hline
\multicolumn{5}{l}{$^a$ All production rates in molecules~s$^{1}$} \\
\end{tabular}
\end{table}

\subsubsection{Reflectance Spectroscopy}\label{section:refl_spectroscopy}

\begin{figure*}
    \centering
    \includegraphics[width=0.47\linewidth]{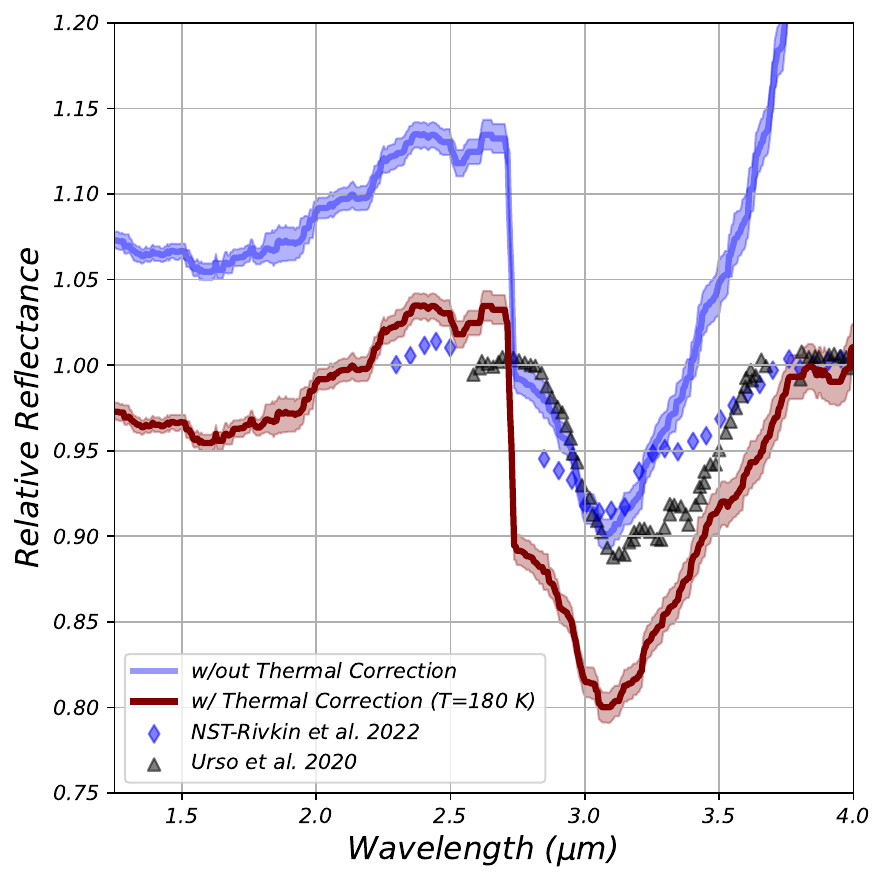}
    \includegraphics[width=0.47\linewidth]{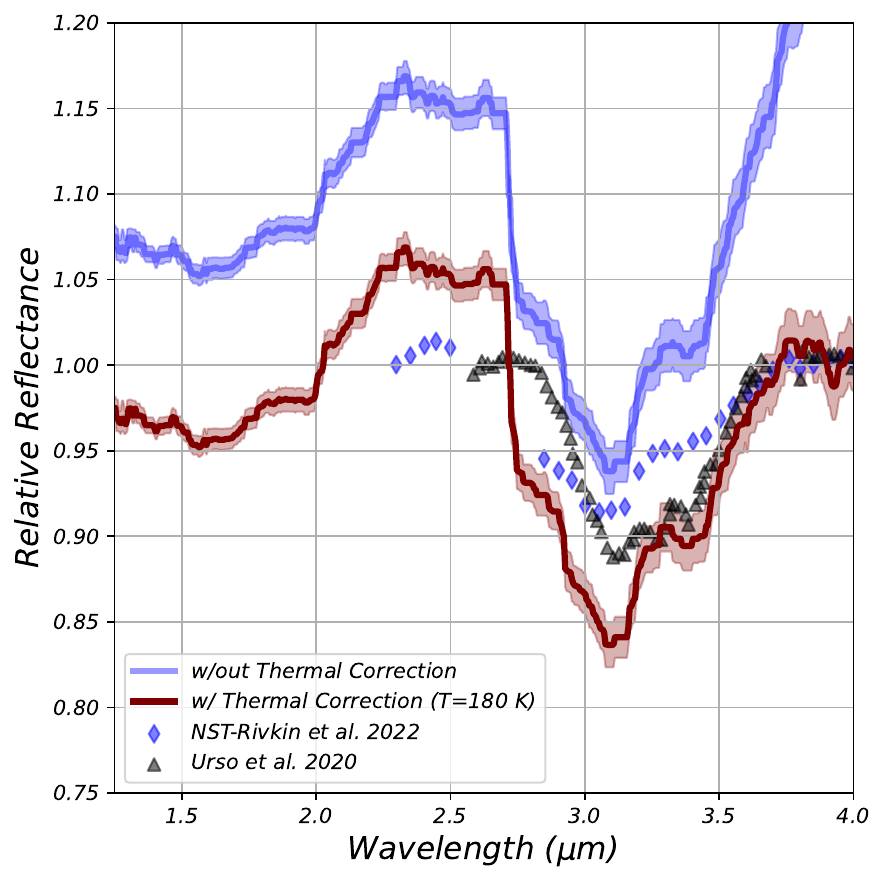}
    \caption{Normalized and linearly corrected reflectance spectra for 133P from UT 2024 June 12 (left) and UT 2024 October 14 (right) with a wide (35 pixel) median filter applied. In both epochs there is a clear asymmetric 3 $\mu$m absorption feature with a band center at 3.1~$\mu$m, much like the non-sharp type low-albedo spectra described in \citet{rivkin2022_3micron}. The band shape and depth beyond 3 $\mu$m in the Cycle 3 observations bears similarity to irradiated water ice and ammonia in experiments carried out by \citet{urso2020_organicirradiation}. }
    \label{fig:Cyc3_NIRSpec_rel_ref}
\end{figure*}

Upon normalizing and correcting our observed spectra for the linear reflectance slope, we see that the 3~$\mu$m absorption band in both epochs has a sharp edge at 2.7~$\mu$m (likely from the OH stretch absorption within phyllosilicates), a maximum absorption at 3.1~$\mu$m, and a shallow redward slope. At 3.3~$\mu$m, we only expect 1\% to 2.6\% of the observed flux to be due to thermal emission contribution, and that the thermal contribution only begins to significantly affect the band shape around 3.5 $\mu$m, so we show the thermally corrected spectrum as well in Figure~\ref{fig:Cyc3_NIRSpec_rel_ref} for clarity.  

The absorption feature present near the H$_2$O emission shown in Figure \ref{fig:133P_H2O_production} is relatively narrow compared to the other features in Figure \ref{fig:Cyc3_NIRSpec_rel_ref}. If we assume that the band center of 2.71 $\mu$m for the second set of observations is more indicative of the true band center due to less H$_2$O emission overlap, that would be consistent with the 2.71-2.72 $\mu$m OH stretch band center seen in CI chondrites due to phyllosilicates like lizardite and chrysotile \citep{takir2013_carbonaceouschondrites}. Given the overlap with the H$_2$O emission it is difficult to definitively characterize this relatively small feature, but we highlight that future JWST observations of 133P at aphelion, when activity is lowest, could provide a clearer view of the surface.

\subsection{Imaging}\label{section:imaging}

\subsubsection{Morphological Analysis - Optical Data}\label{section:morphology_optical}

Examining the composite images of 133P constructed from our optical observations (Figure~\ref{fig:133p_optical_images}), we see unambiguous evidence of activity starting on UT 2024 April 7 when it was at $\nu=352^{\circ}$ and $r_h=2.674$~au (Figure~\ref{fig:133p_optical_images}h; with hints of activity starting as early as UT 2024 March 8, when it was at $\nu=344.8^{\circ}$ and $r_h=2.684$~au).  This activity is seen in the form of a linear tail with a PA aligned with that of the antisolar vector as projected on the sky (which during this time period, coincides with the direction of the comet's negative heliocentric velocity vector).  The appearance of activity over this period of time is consistent with the observed onset of activity for 133P during its 2007 active apparition, when the object first exhibited signs of activity on UT 2007 May 19 when it was at $\nu=349.9^{\circ}$ and $r_h=2.65$~au, after being last observed in an apparently inactive state at $\nu=335.5^{\circ}$ and $r_h=2.68$~au.

Following our initial detection of activity for 133P in 2024, we then continue to see a single tail at constant PA of ${\rm PA}\sim260^{\circ}$, even as the projected PA of the antisolar vector changes by $\sim180^{\circ}$ between UT 2024 July 15 and UT 2024 August 27. On UT 2024 August 27, two tails become clearly visible: one aligned with the PA of the antisolar vector and the second aligned with the PA of the negative heliocentric velocity vector.

This morphology is characteristic of an object displaying sublimation-driven activity \citep[e.g., see][]{hsieh2012_288p}, as it indicates prolonged dust emission based on the simultaneous presence of an antisolar tail comprising smaller particles with short dissipation times (which must have been ejected recently to still be present) and a tail aligned with the negative heliocentric velocity vector comprising larger particles with slower dissipation times (which must have been ejected sufficiently long ago for the tail to reach the observed length).  This conclusion is in agreement with a previous independent dust modeling analysis that also found that 133P's activity to be consistent with sublimation-driven emission \citep{jewitt2014_133p}, as well as 133P's history of recurrent activity during past perihelion passages (see Section~\ref{section:background_133p}), which is widely regarded as a strong indirect indicator of sublimation-driven activity \citep[e.g.,][]{jewitt2024_continuum_comets3}.

\subsubsection{Morphological Analysis - NIRCam Data}\label{section:morphology_nircam}

\begin{figure*}[ht]
    \centering
    \includegraphics[width=0.49\linewidth]{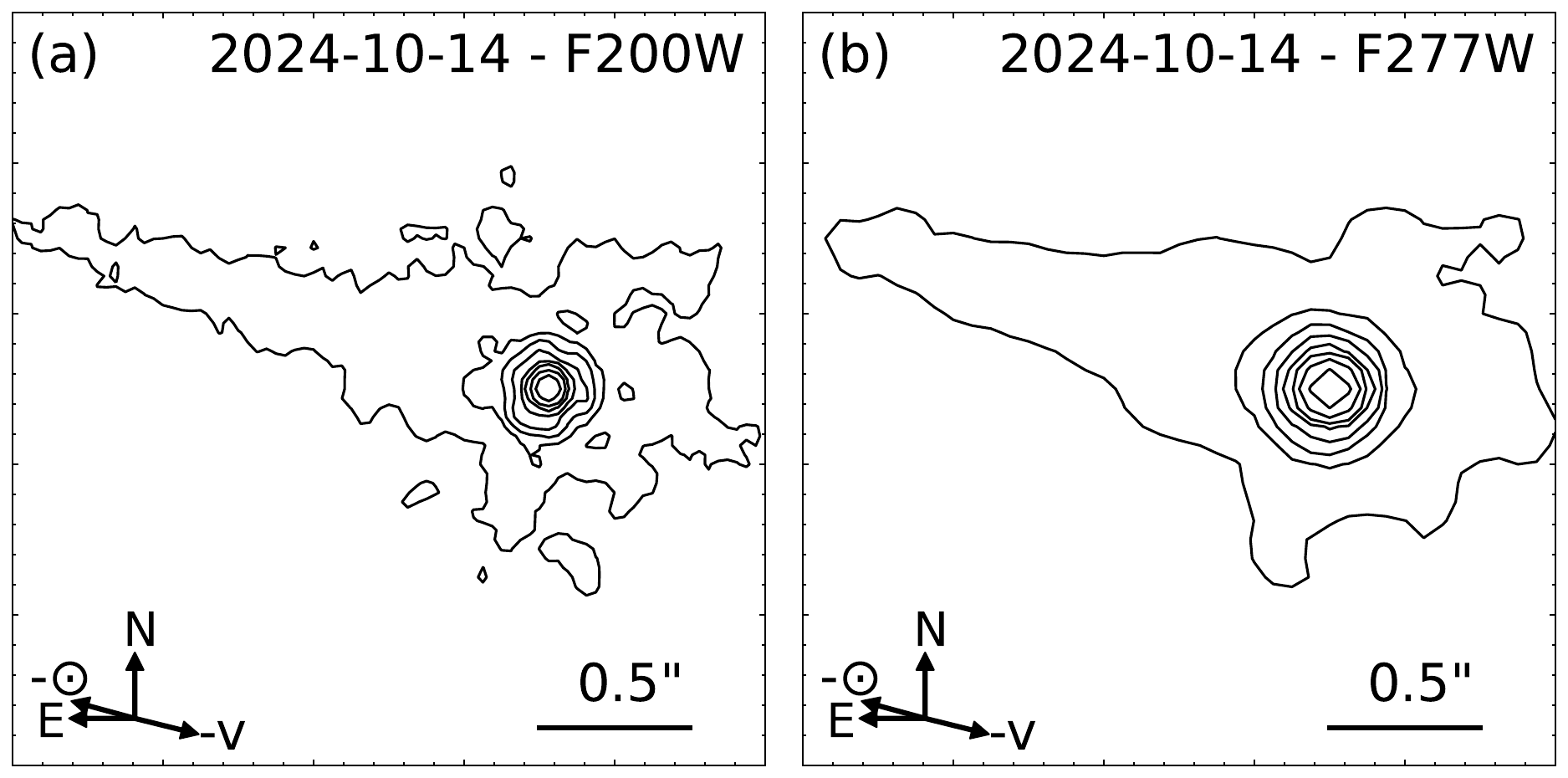}~~~
    \includegraphics[width=0.49\linewidth]{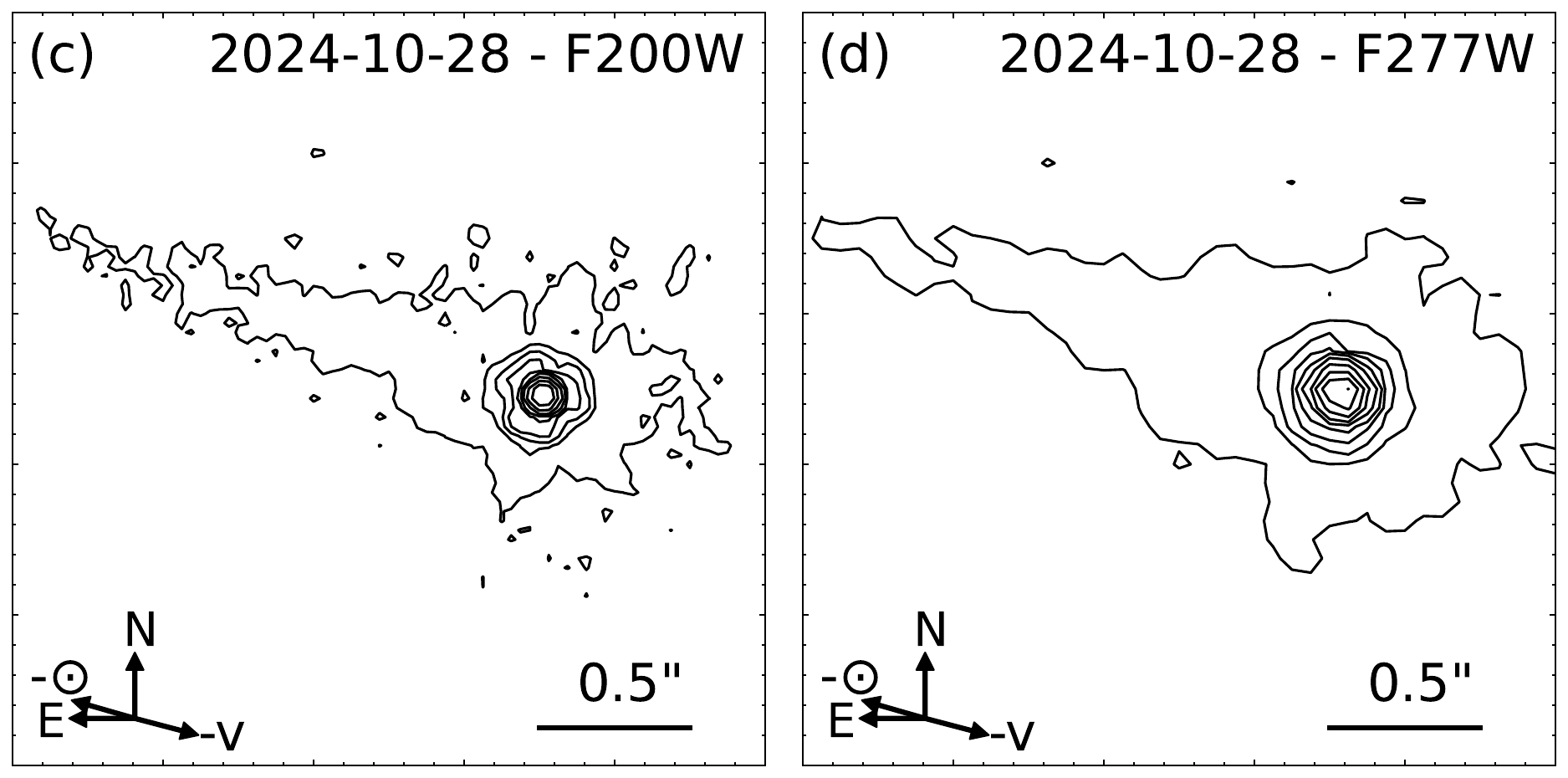}
    \caption{Contour plots of the inner coma of 133P/Elst-Pizarro constructed from (a) F200W and (b) F277W NIRCam median composite images obtained on UT 2024 October 14, and (c) F200W and (d) F277W NIRCam median composite images obtained on UT 2024 October 28, shown in Figure~\ref{fig:nircam_images}.
    Each plot shows 15 logarithmically spaced contour levels ranging from the peak value of each image --- 65.2~MJy/sr in (a), 12.4~MJy/sr in (b), 45.3~MJy/sr in (c), and 10.5~MJy/sr in (d) --- to the background level of $\sim0.01$~MJy/sr (a, b) or $\sim0.03$~MJy/sr (c, d).  A $0\farcs5$ angular scale bar (820 km and 900 km at the distance of the comet on UT 2024 October 14 and UT 2024 October 28, respectively) is shown in each panel, where the orientations of the images and contour plots are the same.}
    \label{fig:nircam_contours}
\end{figure*}

\begin{figure*}[ht]
    \centering
    \includegraphics[width=0.85\linewidth]{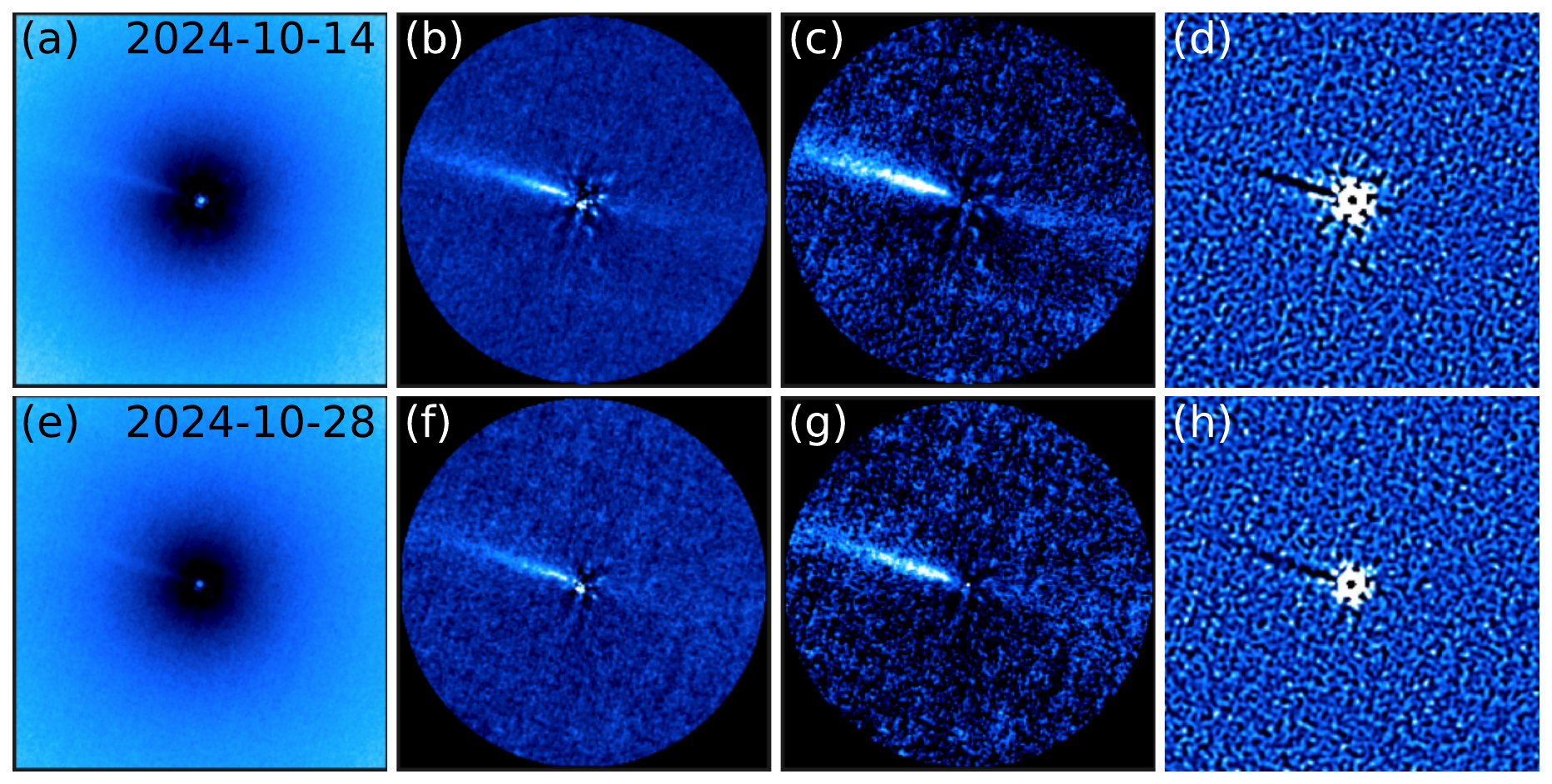}
    \caption{Enhanced F200W composite images of 133P from UT 2024 October 14 (a-d) and UT 2024 October 28 (e-h) using division by a 1/$\rho$ profile (a, e), division by the azimuthal median (b, f), azimuthal renormalization (c, g), and Laplace filtering (d, h).  Each panel is 7$''$$\times$7$''$ in size, with the nucleus of the comet at the center of each image.}
    \label{fig:enhanced_images}
\end{figure*}

NIRCam images and corresponding contour plots of the inner coma of 133P on UT 2024 October 14 and UT 2024 October 28 (Figures~\ref{fig:nircam_images} and \ref{fig:nircam_contours}) show that the point-spread functions (PSFs) on these dates are extremely point-source-like\footnote{see \url{https://jwst-docs.stsci.edu/jwst-near-infrared-camera/ nircam-performance/nircam-point-spread-functions}}, deviating from stellar morphologies and circular symmetry beyond $\rho\sim0\farcs2$ from the photocenter in both the F200W and F277W median composite images from both dates.  Beyond this distance from the photocenter, a visible but low-brightness tail extends in both sets of images towards a position angle of ${\rm PA}\sim75^{\circ}$ East of North, or essentially in the direction of the anti-Solar vector and in the opposite direction of the comet's negative heliocentric velocity vector (Table~\ref{table:jwst_mbc_observations}; Figure~\ref{fig:nircam_images}).

As described in Section~\ref{section:observations}, our two NIRCam visits (for GO 4250 and GO 5551) took place just two weeks apart due to technical problems with carrying out the GO 4250 visit during its originally intended Cycle 2 observing window.  When that visit was missed, we requested and were granted time to re-attempt it in Cycle 3 along with our already planned GO 5551 visit, with the objective of using the two likely closely spaced but non-simultaneous observations to attempt to sample different rotational phases of the nucleus.  With 133P's known rotation period of $P_{\rm rot}=3.471\pm0.001$~hr \citep{hsieh2004_133p} and a total execution time of 26~minutes for our NIRCam sequences (12\% of the rotation period), there was a $\sim$90\% chance of the two observations randomly sampling different portions of the comet's lightcurve.  If we could avoid cases where the same rotational phase range was sampled by both observations, or where the nucleus's subsolar point was close to a rotational pole during the observations, both of which we regarded as being unlikely, we would therefore be able to sample times when different parts of the nucleus were illuminated by the Sun.

If volatile material is not uniformly distributed over 133P's surface, leading to non-uniform dust ejection, we posited that it could be possible to see changes in the near-nucleus morphology of the coma between the two visits, where a lack of observed changes would imply more uniform volatile distribution across the body.  Additionally, given the expected minimal changes in observing and orbital geometry between the two closely temporally spaced visits, any observed morphological changes should be nearly entirely attributable to nucleus rotation.

Examining the observations that were ultimately obtained, we see no immediately apparent changes in near-nucleus morphology between the two visits (see Figures~\ref{fig:nircam_images} and \ref{fig:nircam_contours}).  To investigate further, we apply well-established techniques for enhancing comet images to our composite F200W images (which had both better signal-to-noise ratios and higher spatial resolution than the F277W images) of 133P's near-nucleus coma (see Figure~\ref{fig:nircam_images}).  Specifically, we apply division by a $1/\rho$ profile, division by the azimuthal median, azimuthal renormalization, and Laplace filtering (after $1.5\sigma$ Gaussian smoothing), with results shown in Figure~\ref{fig:enhanced_images}.  The first three sets of enhancements were performed using the online Cometary Coma Image Enhancement Facility\footnote{\url{https://cie.psi.edu/}} \citep{samarasinha2013_cometimageenhancement}, while the last set of enhancements was performed using Gaussian smoothing and Laplace filtering functions in the multidimension image processing package {\tt ndimage} in the {\tt scipy} python library \citep{virtanen2020_scipy}.  Details about each enhancement method are described by \citet{samarasinha2014_cometimageenhancement}, but essentially the first three techniques used here consist of different methods of background estimation and removal to enhance the visibility of faint jet or tail structures, and Laplace filtering attempts to highlight intensity changes in images (i.e., ``edges'') using a convolution filter approximating the second derivative.  
As before, we see no apparent differences in any of the enhanced images for each visit at the spatial resolution (${\textrm{FWHM}}\sim0\farcs07$) of the images. 

Although we do not note any significant changes in enhanced images of 133P from the two NIRCam visits, we note that in the images enhanced using division by the azimuthal median (Figures~\ref{fig:enhanced_images}b,f) and azimuthal renormalization (Figures~\ref{fig:enhanced_images}c,g), there appears to be a faint approximately sunward feature opposing the brighter antisolar tail.  This appears to be the remnant of an opposing tail (seen more clearly in optical images from UT 2024 August 27 through approximately UT 2024 October 28; Figure~\ref{fig:133p_optical_images}s-y) consisting of older, larger particles than are in the antisolar tail \citep[e.g., as was seen for 288P;][]{hsieh2012_288p}, which is a strong sign of sublimation-driven activity (see Section~\ref{section:morphology_optical}).

Given the timing of the actual observations (Table~\ref{table:jwst_mbc_observations}) and assigning a rotational phase of $\phi=0$ to the start of the first NIRCam sequence (on UT 2024 October 14), we find that the relative rotational phase at the start of the second sequence was actually at $\phi=0.89\pm0.08$, meaning that, within uncertainties, both observations may have sampled the same portion of 133P's surface.  Additionally, we note that while the  relative simplicity of the comet's morphology (i.e., a single straight, narrow tail, with no definitive evidence of more complex coma structure that could imply the presence of collimated dust emission from localized active sites) and lack of changes between the two NIRCam visits is consistent with uniform distribution of sublimating volatile material across the object's surface, it is also consistent simply with the slow ejection of large dust particles (see Section~\ref{section:dust_production}).

In the latter case, the motion of ejected dust particles is dominated by solar gravity and solar radiation pressure, meaning that, at the spatial scales we are able to probe with NIRCam observations, the coma morphology carries little to no information about initial dust ejection trajectories and distribution that could hint at the comet's active area distribution.  Given these considerations, we find our effort to characterize the likely active area distribution on 133P by analyzing its dust morphology in NIRCam observations to be inconclusive.

\subsection{Photometry}\label{section:photometry}

\subsubsection{NIRCam Photometry}\label{section:nircam_photometry}

Photometric measurements of our NIRCam data were carried out following a \jwst{} Data Analysis Tool Jupyter notebook\footnote{\url{https://spacetelescope.github.io/jdat_notebooks/index.html}} provided by the Space Telescope Science Institute (STScI) to facilitate post-pipeline analysis of \jwst{} data.
Measurements of the four NIRCam observations of 133P obtained on UT 2024 October 14 result in measured average AB magnitudes of $m_{\rm F200W}=(20.98\pm0.14)$~mag and $m_{\rm F277W}=(21.49\pm0.11)$~mag, while we find  average magnitudes of $m_{\rm F200W}=(21.38\pm0.08)$~mag and $m_{\rm F277W}=(21.91\pm0.10)$~mag for the four NIRCam observations obtained on UT 2024 October 28.  Listed uncertainties correspond to the standard deviation of the individual photometric measurements included in each average and likely include large contributions from the nucleus's known large rotational lightcurve amplitude and relatively short rotation period, in addition to ordinary measurement uncertainties.

These results yield ${\rm F200W}-{\rm F277W}$ colors of $-0.51\pm0.18$ on UT 2024 October 14 and $-0.53\pm0.13$ on UT 2024 October 28.  While nominally consistent with all ${\rm F200W}-{\rm F277W}$ colors measured for previously observed MBCs within $2\sigma$, these colors are nonetheless more similar to those measured for 457P \citep[$-0.63\pm0.08$;][]{noonan2025_jwst457p} than those measured for 238P \citep[$-0.38\pm0.05$;][]{kelley2023_jwst238p} and 358P 
\citep[$-0.29\pm0.05$;][]{hsieh2025_358p}.
The slightly redder colors of 238P and 358P could be plausibly attributed to the smaller dust-to-gas ratios (where the F277W bandpass includes the 2.7~$\mu$m water emission band) of the two objects at the time of their observations compared to 133P and 457P.

\subsubsection{Optical Photometry}\label{section:optical_photometry}

For our optical data, photometry measurements of 133P and at least one background reference star were performed using {\tt IRAF} \citep{tody1986_iraf,tody1993_iraf,fitzpatrick2024_iraf} and {\tt pyraf} software.
Photometry of the target was performed using circular apertures with angular radii listed in Table~\ref{table:ground_photometry_133p}. These aperture radii were selected using curve of growth analyses to select radii large enough to encompass the distance at which the surface brightness profile of the object reached the level of the background sky in each image, but small enough to avoid inclusion of flux from nearby background sources as well as excessive background sky noise. Background statistics for these photometry measurements were measured in nearby but non-adjacent regions of blank sky to avoid contamination from the object's own dust or flux from nearby background sources (where we achieve single-exposure signal-to-noise ratios of ${\rm S/N}>20$ for all observations of our target).  
When composite images (see Section~\ref{section:optical_data_processing}) showed that background sources were within the optimal photometry apertures determined from curve-of-growth analyses, indicating that photometry was likely to be unreliable, photometric measurements of those detections were rejected.

Absolute photometric calibration was performed using field star magnitudes from the Asteroid Terrestrial-impact Last Alert System \citep[ATLAS;][]{tonry2011_atlas,tonry2018_atlas} Refcat2 all-sky stellar reference catalog \citep{tonry2018_refcat}, where conversion of non-SDSS photometry to magnitudes in the SDSS system was accomplished as needed using transformations derived by \citet{tonry2012_ps1} and by R.~Lupton\footnote{{\url{http://www.sdss.org/}}}. 
Similar to photometry of the target object, photometry of reference stars was obtained by measuring net fluxes within circular apertures with sizes chosen using curve-of-growth analyses of representative stars, with background sampled from surrounding circular annuli.
We aimed to use $5-30$ well-isolated reference stars (i.e., field stars with no other neighboring sources within the photometry aperture used for those data, and ideally, within the annuli used to measure sky background as well) for photometric calibration where possible.
In some cases, however, only a few suitable reference stars, or even just one, were available due to the small margin between the limiting magnitude of the Refcat2 catalog and the saturation limit of many of our observations.

To characterize the comet's intrinsic brightness evolution, calibrated photometry were normalized to heliocentric and geocentric distances of $r_h=\Delta=1$~au, producing reduced magnitudes, and then to a Solar phase angle of $\alpha=0^{\circ}$ to produce equivalent absolute magnitudes. Reduced magnitudes were computed by applying a correction of $-5\log (r_h\Delta)$ to the measured apparent magnitudes. Absolute magnitudes were then computed by determining the nucleus's reduced magnitude at the observed phase angle ($\alpha$) using the IAU phase function parameters computed by \citet{hsieh2023_mbcnuclei}, subtracting the nucleus's reduced magnitude from the total measured reduced magnitude of the comet (i.e., leaving the reduced magnitude of the dust), correcting the reduced magnitude of the dust to $\alpha=0^{\circ}$ using the Schleicher-Marcus phase function\footnote{\url{https://asteroid.lowell.edu/comet/dustphase.html}} \citep[sometimes also referred to as the Halley-Marcus phase function;][]{schleicher2011_sw3,schleicher1998_halley,marcus2007_cometphasefunction} to obtain the effective absolute magnitude of the dust, $m_d(1,1,0)$, and finally adding back the absolute magnitude of the nucleus ($H_V$).  These absolute magnitude results (also converted to $V$-band, assuming solar colors) are shown in Table~\ref{table:ground_photometry_133p}.

As part of these calculations, we are also able to compute the ratio of scattering cross-sections of the dust and the nucleus, $A_d/A_n$, where
\begin{equation}
    {A_d\over A_n} = 10^{0.4\left[H_V-m_d(1,1,0)\right]}
\end{equation}
where these values are also shown in Table~\ref{table:ground_photometry_133p}.
We note that these calculations assume that the dust coma is optically thin. 
These calculations and others described in subsequent sections use the {\tt uncertainties} {\tt python} package for the calculation and propagation of uncertainties\footnote{\url{http://pythonhosted.org/uncertainties/}}.

\subsubsection{$Af\rho$ from Optical Data}\label{section:afrho_optical}

To quantitatively characterize 133P's dust content during its 2024 active apparition and use it to place our derived water production rates (Section~\ref{section:H2O_analysis}) in context, we compute the quantity $A(0^{\circ})f\rho$, hereafter $Af\rho$, for each set of our optical observations.  $Af\rho$ is commonly used as a proxy for dust production rate that can be used to compare dust production activity levels derived from nucleus-subtracted and phase-function-corrected photometric measurements made of cometary coma observed at different times and under different conditions \citep{ahearn1984_bowell}.  It is given by
\begin{equation}
    Af\rho = {(2r_h\Delta)^2\over\rho} 10^{0.4[m_{\odot}-m_d(r_h,\Delta,0)]}
    \label{eqn:afrho}
\end{equation}
where $r_h$ is in au, $\Delta$ is in cm, and the physical radius of the photometry aperture used to measure the magnitude of the comet at the distance of the comet, $\rho$, is in cm. The apparent magnitude of the Sun, $m_{\odot}$, at $\Delta=1$~au in the same filter used to observe the comet is computed using $m_{\odot,V}=-26.71\pm0.03$ from \citet{hardorp1980_sun3}, and $m_d(r_h,\Delta,0)$ is the phase-angle-corrected (to $\alpha=0^{\circ}$) apparent magnitude of the dust with the flux contribution of the nucleus subtracted from the measured total magnitude.

\begin{figure}[htb]
    \centering
    \includegraphics[width=1.0\linewidth]{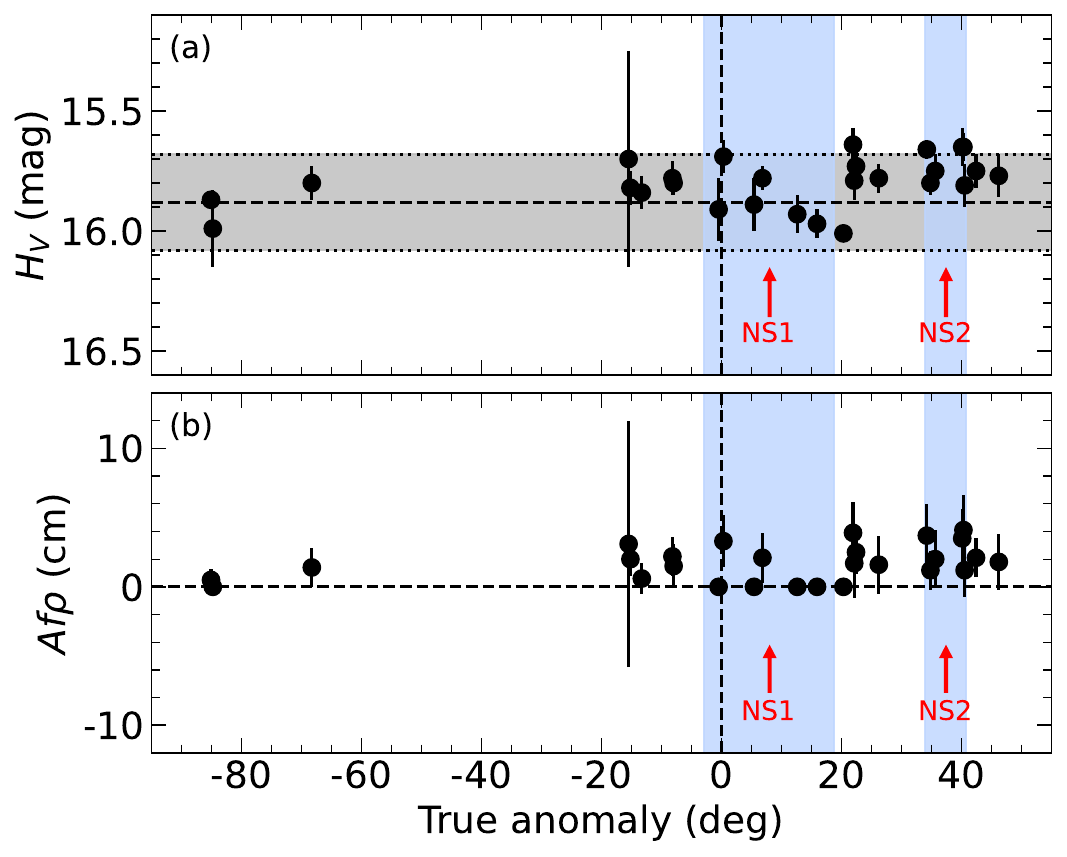}
    \caption{Plots of (a) equivalent total absolute $V$-band magnitude and (b) $Af\rho$ measured for 133P as functions of true anomaly.  In panel (a), the dashed horizontal line shows the absolute $V$-band magnitude of 133P's inactive nucleus of $H_V=15.88$~mag previously reported by \citet{hsieh2023_mbcnuclei}, and the gray-shaded area bounded by horizontal dotted lines indicates the range of potential brightness variations allowed by the assumption of $\Delta m=0.2$~mag. In panel (b), the dashed horizontal line shows $Af\rho=0$, i.e., for the absence of activity.  In each panel, the true anomalies of the first and second NIRSpec observations are marked by ``NS1'' and ``NS2'', respectively, and the blue shaded regions indicate the ranges around each NIRSpec observation that are used to calculate the effective equivalent $Af\rho$ value for the target at the time of those observations.}
    \label{fig:absmag_afrho}
\end{figure}

$Af\rho$ is nominally independent of $\rho$ for a spherically symmetric, steady-state coma with a line-of-sight column density that scales with $\rho^{-1}$ (which results from a three-dimensional density profile that scales with $\rho^{-2}$) and no production or destruction of dust grains in the coma.  As noted in \citet{hsieh2025_358p}, however, while asymmetric dust ejection (e.g., in the form of jets or fans) should still produce an observed $\rho^{-1}$ radial surface brightness profile \citep{protopapa2014_103p}, maintaining the independence of $Af\rho$ of aperture size, steeper surface brightness profiles can result at distance scales where coma morphology becomes dominated by radiation pressure effects \citep{jewitt1987_cometsbps,fink2012_afrho}, which, in practice, is the regime in which all of our ground-based photometry is done, given the geocentric distance of 133P at the time of our observations and prevailing seeing conditions at those times \citep[Table~\ref{table:ground_observations_133p}; also see][]{hsieh2025_358p}.  Such steeper surface brightness profiles result in $Af\rho$ measurements that are not independent of $\rho$, but instead decrease with increasing $\rho$. 

\begin{figure}[ht]
    \centering
    \includegraphics[width=1.0\linewidth]{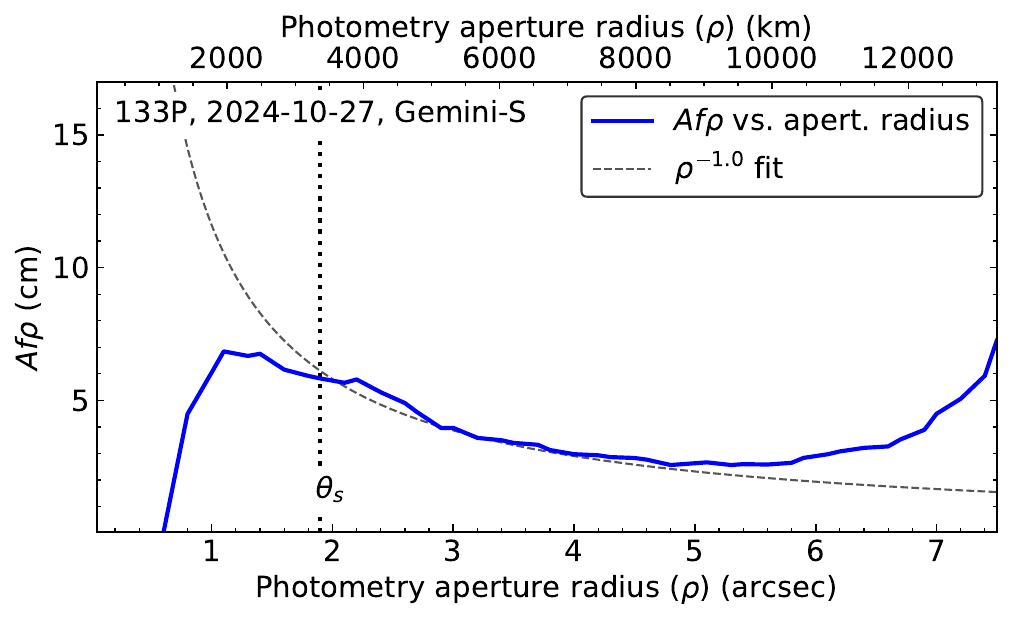}
    \caption{Plot of $Af\rho$ as a function of photometry aperture radius (solid blue line) in terms of arcseconds projected on the sky (bottom $x$-axis labels) and km at the distance of the comet (top $x$-axis labels) as measured for a composite image of 133P constructed from data obtained on UT 2024 October 27 by Gemini South, with a $\rho^{-1.0}$ power law (dashed black line) shown for reference.  The seeing ($\theta_s$) from Table~\ref{table:ground_observations_133p} is marked by a dotted vertical line.}
    \label{fig:afrho_profile}
\end{figure}

Given the above considerations, in order to facilitate comparisons to previously published $Af\rho$ measurements, we follow the method used by \citet{hsieh2025_358p} of computing $Af\rho$.  We first identify the optimal photometry aperture for the particular data being measured (which ranged from $1\farcs6$ to $5\farcs5$ for the data we report here; see Table~\ref{table:ground_photometry_133p} and Section~\ref{section:optical_data_processing}), and compute $Af\rho$ from photometry using that aperture size.  We then extrapolate that $Af\rho$ measurement to a physical aperture size of $\rho=5000$~km at the distance of the comet (equivalent to $2\farcs2$ to $4\farcs1$ for the data we report here; listed in Table~\ref{table:ground_photometry_133p}).
This final physical aperture size is chosen so that our reported measurements are physically consistent across our observations, but is also comparable to aperture sizes typically used for other published $Af\rho$ measurements.  

For this extrapolation, we examine a relatively high-quality $Af\rho$ profile plotted as a function of $\rho$ that we measure for detections (see Figure~\ref{fig:133p_optical_images}x) of 133P obtained on UT 2024 October 27 by Gemini South.  We find a good fit to the measured profile using a power law that scales with $\rho^{-1.0}$  (Figure~\ref{fig:afrho_profile}), and thus,  given the difficulty of conducting independent fits for detections with contamination from nearby background sources, adopt this as the assumed power law form for the $Af\rho$ profiles for each of our observations.

Following \citet{hsieh2025_358p}, to facilitate comparisons with other published work, we apply uncertainties for our $Af\rho$ values corresponding to aperture size ranges of $\rho=(5000\pm2000)$~km and add these in quadrature to the uncertainties arising from the underlying photometry.

These calculations were conducted for all sets of observations for which the measured flux-averaged total brightness of 133P was brighter than the expected brightness of its inactive nucleus at the midpoint of its rotational lightcurve based on the absolute magnitude of $H_V=15.88\pm0.04$ reported by \citet{hsieh2023_mbcnuclei}.  For other measurements where the measured flux-averaged total brightness of 133P was fainter than the expected brightness of its inactive nucleus, we assign placeholder values of 0~cm for $Af\rho$ (since $Af\rho$ is mathematically undefined in these cases; see Equation~\ref{eqn:afrho}).  These $Af\rho$ values and the $H_V$ values used to derive them are plotted as a function of true anomaly in Figure~\ref{fig:absmag_afrho}.

Notably, instances when $Af\rho$ is undefined due to 133P being fainter than predicted from the inactive nucleus's measured absolute magnitude occurred several times when the object was clearly active (see Table~\ref{table:ground_photometry_133p}, and Figures~\ref{fig:absmag_afrho}a and \ref{fig:133p_optical_images}).  
They can be explained as a result of the nucleus's substantial elongation \citep[producing rotational lightcurve amplitudes of up to $\Delta m\sim0.2$~mag; e.g., see][]{hsieh2004_133p}, and the object's extremely weak dust production at the time (where the estimated photometric enhancement from ejected dust around the time of the first NIRSpec observations is just $\sim0.03$~mag; see below).  Negative rotational brightness variations can therefore easily exceed the small photometric enhancement provided by ejected dust during this time, causing the object to appear fainter than the expected rotationally averaged brightness of the inactive nucleus, despite having visible activity.

Due to the multiple instances where $Af\rho$ was undefined around the time of our UT 2024 June 12 NIRSpec observations, instead of averaging the remaining valid $Af\rho$ values from that period, we compute $Af\rho$ for that set of NIRSpec observations based on the midpoint absolute magnitude value of $H_{V,{\rm mid}}\sim15.85$ (i.e., halfway between the minimum and maximum values) over a three month period centered on those observations (i.e., 45 days before and after, or from UT 2024 April 28 to UT 2024 July 27).  This period is shown by the earlier blue-shaded region in Figure~\ref{fig:absmag_afrho} and also indicated in Table~\ref{table:ground_photometry_133p}). For the photometry aperture size and other observing geometry parameters used for the $Af\rho$ calculation, we use parameters associated with the Gemini-S observations obtained on UT 2024 June 7 that were obtained closest in time to the NIRSpec observations.

Meanwhile, for our UT 2024 October 14 NIRSpec observations, we use the median value of $Af\rho$ measurements obtained 15 days before and after those observations (i.e., between UT 2024 September 29 and UT 2024 October 29; shown by the later blue-shaded region in Figure~\ref{fig:absmag_afrho} and indicated in Table~\ref{table:ground_photometry_133p}) as the effective $Af\rho$ value at the time of those observations. We adopt the standard deviation of the $Af\rho$ measurements obtained during that period as the uncertainty on the values for both periods.

Using the approaches described above, we find $Af\rho=(0.8\pm1.5)$~cm and $Af\rho=(2.8\pm1.5)$~cm at the times of our first and second sets of NIRSpec observations, respectively. While these values are formally consistent with one another within their assigned 1-$\sigma$ uncertainties, suggesting that there is no statistically meaningful difference in dust production between the two sets of observations, we note that the mean absolute magnitudes from Table~\ref{table:ground_photometry_133p} (computed in flux space) for the photometry measured during the first and second sets of observations are $m_{V,{\rm avg}}(1,1,0) =(15.84\pm0.03)$~mag and $m_{V,{\rm avg}}(1,1,0)=(15.71\pm0.02)$~mag, respectively.  Meanwhile, the median absolute magnitudes for the same time periods (adopting the standard deviation of the measurements for each period as 1-$\sigma$ uncertainties) are $m_{V,{\rm med}}(1,1,0)=15.90\pm0.10$~mag and $m_{V,{\rm med}}(1,1,0)=(15.71\pm0.07)$~mag.  Therefore, by either metric, we find a brightening of the near-nucleus coma between the two observing period that is meaningful to at least the 1$\sigma$ level, implying a similarly meaningful increase in the measured near-nucleus dust mass.

\subsubsection{$Af\rho$ from NIRCam Data}\label{section:afrho_nircam}

For reference, we also compute $Af\rho$ from the UT 2024 October 14 NIRCam observations that were obtained during the same visit as the second set of NIRSpec observations.  In ground-based optical data obtained between UT 2024 September 30 and UT 2024 October 28 (bracketing the October \jwst{} observations), the apparent $V$-band brightness of the dust comprised $\sim0.2$ of the total measured $V$-band brightness of the comet. Due to the different assumed phase function behavior of the nucleus and the dust discussed in Section~\ref{section:optical_data_processing}, this fraction differs from the one implied by the $A_d/A_n$ value listed in Table~\ref{table:ground_photometry_133p}, which was computed from absolute magnitudes.

The total measured magnitudes of the comet in the F200W and F277W data, respectively, are $m_{\rm F200W}=(20.98\pm0.14)$~mag and $m_{\rm F277W}=(21.49\pm0.11)$~mag (Section~\ref{section:nircam_photometry}).
Assuming the same contribution from the dust to the total brightness of the comet at near-infrared wavelengths as in optical data, we obtain apparent dust magnitudes of $m_{d,{\rm F200W}}\sim22.58$~mag and $m_{d,{\rm F277W}}\sim23.09$~mag, and phase-angle-corrected dust magnitudes of $m_{d,{\rm F200W}}(0^{\circ})\sim21.80$~mag and $m_{d,{\rm F277W}}(0^{\circ})\sim22.31$~mag, assuming the Schleicher-Marcus phase function (see Section~\ref{section:optical_data_processing}).
Using Equation~\ref{eqn:afrho}, $m_{\odot}=-26.64$~mag for F200W, and $m_{\odot}=-26.03$~mag for F277W, we then compute $Af\rho_{\rm F200W}\sim29.8$~cm and $Af\rho_{\rm F277W}\sim32.6$~cm.  Using the same scaling factor as used by \citet{kelley2023_jwst238p}, these results correspond to $Af\rho\sim24$~cm at 0.7~$\mu$m.  Then extrapolating to $\rho=5000$~km assuming a $\rho^{-1.0}$ power law (see above), we finally find $Af\rho_{\rm 5000km}\sim2.2$~cm, which is in reasonable agreement within uncertainties with results from ground-based optical data obtained around the same time (see Section~\ref{section:afrho_optical}).

\subsubsection{Dust Production Rate Characterization}\label{section:dust_production}

Following the method described by \citet{kim2022_p2020o1} and \citet{noonan2025_jwst457p}, we can estimate the dust production rate of 133P at the time of our \jwst{} observations.  Following Equation~\ref{eqn:afrho}, we compute the absolute magnitude of the dust coma corresponding to the $Af\rho$ values computed in Section~\ref{section:afrho_optical} using
\begin{equation}
    m_{d,V}(1,1,0)=m_{\odot} - 2.5\log{(Af\rho)\rho\over (2r_h\Delta_{\rm cm})^2} - 5\log(r_h\Delta_{\rm au})
\end{equation}
where $r_h$ is in au, $\Delta_{\rm cm}$ and $\Delta_{\rm au}$ are in cm and au, respectively, and $\rho$ is in cm.
Using $(Af\rho)_1=(0.8\pm1.5)$~cm and $(Af\rho)_2=(2.8\pm1.5)$~cm for our first and second sets of NIRSpec observations, respectively (Section~\ref{section:afrho_nircam}), we find
$m_{d,V,1}(1,1,0)=19.2\pm2.0$ and
$m_{d,V,2}(1,1,0)=17.8\pm0.6$.

The total cross section of the dust (in km$^2$) can then be computed using
\begin{equation}
    C_d = \frac{2.24 \times 10 ^{16} \pi}{p_V} 10^{0.4[m_{\odot,V}-m_{d,V}(1,1,0)]}
\label{eqn:cross_section}
\end{equation}
\citep{kim2022_p2020o1}, where we assume a geometric albedo of $p_V=0.05$, and use $m_{\odot,V}=-26.71\pm0.03$ for the $V$-band apparent magnitude of the Sun \citep{hardorp1980_sun3}. We find total dust scattering cross sections of
$C_{d,1}=(0.6\pm1.2)$~km$^2$ and
$C_{d,2}=(2.2\pm1.2)$~km$^2$.

The equivalent dust mass, $M_d$, corresponding to these scattering cross-sections can be estimated using
\begin{equation}
    M_d={4\over3}C_d{\bar a_d}\rho_d
    \label{eqn:dust_mass}
\end{equation}
where ${\bar a_d}$ is the effective mean dust grain radius and $\rho_d$ is the dust grain bulk density, assuming that the coma is optically thin.  Assuming ${\bar a_d}\sim3$~mm \citep[from][]{jewitt2014_133p} and $\rho_d= 2500$~kg~m$^{-3}$ \citep[similar to that of CI and CM carbonaceous chondrites;][]{britt2002_astdensities_ast3}, we find dust coma masses of $M_{d,1}=(0.6\pm1.2)\times10^{7}$~kg and $M_{d,2}=(2.2\pm1.2)\times10^{7}$~kg.

Analysis of \hst{} observations obtained during 133P's 2013 active apparition indicated characteristic dust velocities of $v_d\sim1.8a_{d}^{-1/2}$~m~s$^{-1}$, where $a_{d}$ is the particle radius in $\mu$m.  Using $v_d=0.03$~m~s$^{-1}$ for $\bar a_d=3$~mm dust grains, we find a minimum aperture-crossing time of $\tau_{\rm ap}\sim(1.5\times10^{8})$~s for $\rho=5000$~km apertures (assuming fully transverse motion as projected on the sky; ejected dust with line-of-sight velocity components would spend even longer in the aperture).  This crossing time is equivalent to 4.75~yr, which is much longer than the timescales over which morphology changes have been observed in the past \citep[e.g.,][]{hsieh2004_133p,hsieh2010_133p}. As such, it is more reasonable to assume that, after dust particles are lofted from the nucleus at these extremely slow ejection velocities, their motion is then largely dominated by solar gravity and radiation pressure.  In this case, the residence time in the photometry aperture is given instead by $\tau_{\rm ap}\sim(2\rho/\beta_d g_{\odot})^{1/2}$ \citep{kim2022_p2020o1}, where $g_{\odot}=10^{-3}$~m~s$^{-2}$ is the solar gravitational acceleration and we adopt an effective $\beta_d$ parameter value of $\bar \beta_d=3\times10^{-4}$ from \citet{jewitt2014_133p}.  This calculation produces a significantly shorter minimum aperture crossing time of $\tau_{\rm ap}\sim6\times10^6$~s ($\sim70$ days, which is more in line with the timescale of morphological changes observed for 133P in the past),
implying dust production rates of $Q_{d,1}=(1.0\pm2.1)$~kg~s$^{-1}$ and $Q_{d,2}=(3.8\pm2.1)$~kg~s$^{-1}$.

\subsection{Dust-to-Gas Ratios}\label{section:dust_to_gas}

Using $Q_{\rm H_2O}$ and $Af\rho$ values computed in Sections~\ref{section:H2O_analysis} and \ref{section:afrho_optical}, respectively,
we compute dust-to-gas ratios of $\log(Af\rho/Q_{\rm H_2O})=-25.4\pm0.8$ and $\log(Af\rho/Q_{\rm H_2O})=-24.7\pm0.3$ for our first and second NIRSpec observations of 133P, respectively.  133P's coma therefore appears significantly more gas-dominated during the first NIRSpec visit, with $Af\rho/Q_{\rm H_2O}$ nominally increasing by a factor of 5 between the two visits.  For reference, this change in the dust-to-gas ratio is comprised of a nominal 25\% decline in the measured $Q_{\rm H_2O}$ rate (Section~\ref{section:H2O_analysis}) and a factor of 3.5 increase in estimated $Af\rho$ values (Section~\ref{section:afrho_optical}).

Despite the apparently large increase in the gas-to-dust ratio between our two NIRSpec visits, the computed $\log(Af\rho/Q_{\rm H_2O})$ values for the two visits are consistent within uncertainties (similar to our water production measurement results; Section~\ref{section:H2O_analysis}), where the value computed for the first visit has a particularly large uncertainty that is more than twice as large (in log space) as the corresponding uncertainty for the second visit.  This large uncertainty in the first $\log(Af\rho/Q_{\rm H_2O})$ measurement appears to be significantly impacted by the large assigned uncertainty of 1.5~cm on the $Af\rho$ estimate for that visit (which was based on the standard deviation of $Af\rho$ values computed for the second visit; Section~\ref{section:afrho_optical}), where the uncertainty is almost twice that of the nominal value ($Af\rho=0.8$~cm). For comparison, the same uncertainty is about half of the nominal value ($Af\rho=2.8$~cm) used for the second visit.

If we therefore exclude the $\log(Af\rho/Q_{\rm H_2O})$ value computed for our first NIRSpec observations of 133P as being poorly constrained, we are left with a $\log(Af\rho/Q_{\rm H_2O})$ value for the second visit that is quite consistent with results found for 238P and 358P \citep[$\log(Af\rho/Q_{\rm H_2O})=-24.4\pm0.2$ and $\log(Af\rho/Q_{\rm H_2O})=-24.8\pm0.2$, respectively;][]{kelley2023_jwst238p,hsieh2025_358p}.  Combining the results for these three objects, we find an average value of $\log(Af\rho/Q_{\rm H_2O})=-24.6\pm0.2$ (where the standard deviation of the sample is assigned as the uncertainty).

Meanwhile, based on the water production rates measured in Section~\ref{section:H2O_analysis} and dust production rates estimated in Section~\ref{section:dust_production}, we derive dust-to-gas production rate ratios by mass of $Q_d/Q_{\rm H_2O}=1.7\pm3.5$ and $Q_d/Q_{\rm H_2O}=9.5\pm5.8$ at the times of our first and second NIRSpec visits (Table~\ref{table:dust_vs_water}), respectively. 
The dust production rates used to compute these ratios were derived from $Af\rho$ estimates, however, and as such, given the discussion above, we regard the second measurement to be more reliable.
However, this leaves a measurement that is significantly higher (even within uncertainties) than the corresponding values of $Q_d/Q_{\rm H_2O}\sim0.8$ and $Q_d/Q_{\rm H_2O}\sim0.5$ found for 238P and 358P, respectively \citep{kelley2023_jwst238p,hsieh2025_358p}.
We discuss this discrepancy further in Section~\ref{section:comparison_dusttogas}.

\setlength{\tabcolsep}{8pt}
\setlength{\extrarowheight}{0em}
\begin{table*}[htb]
\caption{Main-Belt Comet \jwst{} NIRSpec Observations}
\centering
\smallskip
\footnotesize
\begin{tabular}{cccccccrrr}
\hline\hline
\multicolumn{1}{c}{Target}
 & \multicolumn{1}{c}{$a$$^a$}
 & \multicolumn{1}{c}{$e$$^b$}
 & \multicolumn{1}{c}{$i$$^c$}
 & \multicolumn{1}{c}{$H_V$$^d$}
 & \multicolumn{1}{c}{$r_n$$^e$}
 & \multicolumn{1}{c}{Obs.\ Date$^f$}
 & \multicolumn{1}{c}{$\nu$$^g$}
 & \multicolumn{1}{c}{$r_h$$^h$}
 & \multicolumn{1}{c}{$\Delta t_q$$^i$}
 \\
\hline
133P & 3.163 & 0.156 & 1.390 & 15.88$\pm$0.04 & 2.0$\pm$0.5 & 2024 Jun 12 &  8.0 & 2.674 & $+$33 \\ % obs=2460473.8 q=2460440.8
...  & ... & ... & ... & ... & ... & 2024 Oct 14 & 37.4 & 2.747 & $+$157 \\ % obs=2460597.8 q=2460440.8
238P & 3.166 & 0.252 & 1.264 & 20.5$\pm$0.1 & 0.24$\pm$0.06 & 2022 Sep 08 & 28.3 & 2.428 & +95  \\
358P & 3.147 & 0.239 & 11.060 & 20.2$\pm$0.3 & 0.3$\pm$0.1 & 2024 Jan 08 & 17.4 & 2.416 & +59  \\
457P & 2.646 & 0.120 & 5.222 & $18.69\pm0.06$ & $0.56\pm0.06$ & 2024 Sep 20 &  9.2 & 2.335 & +31  \\
\hline
\hline
\multicolumn{10}{l}{$^a$ Semimajor axis, in au.} \\
\multicolumn{10}{l}{$^b$ Eccentricity.} \\
\multicolumn{10}{l}{$^c$ Inclination, in degrees.} \\
\multicolumn{10}{l}{$^d$ $V$-band absolute magnitude, from \citet{hsieh2023_mbcnuclei}.} \\
\multicolumn{10}{l}{$^e$ Effective circular nucleus cross-sectional radius, in km, from \citet{hsieh2023_mbcnuclei}.} \\
\multicolumn{10}{l}{$^f$ UT observation date by NIRSpec on \jwst{}.} \\
\multicolumn{10}{l}{$^g$ True anomaly at the time of observation by NIRSpec, in degrees, from JPL Horizons.} \\
\multicolumn{10}{l}{$^h$ Heliocentric distance at the time of observation by NIRSpec, in au, from JPL Horizons.} \\
\multicolumn{10}{l}{$^i$ Time relative to perihelion (positive values indicating time after perihelion), in days.} \\
\end{tabular}
\label{table:jwst_mbc_observations_summary}
\end{table*}

\section{Discussion}

\subsection{Water Production Evolution}\label{section:water_evolution}

The set of observations of 133P reported in this work are unique in that they represent the first time the water production rate of a MBC has been measured by \jwst{} at two points in its orbit, enabling us to measure for the first time how this parameter varies as a function of heliocentric distance and/or orbit position for the same object during the same active apparition.  As discussed in Section~\ref{section:H2O_analysis}, we see a nominal 25\% decline in $Q_{\rm H_2O}$ between our two NIRSpec visits, although within uncertainties, water production may also have remained constant or even slightly increased between visits.
Regardless of the exact activity strength evolution between these visits, however, we find the simple yet significant result that 133P produced measurable amounts of water vapor in observations four months apart, indicating that its 2024 activity was characterized by prolonged gas (and therefore dust) emission. This result is consistent with conclusions inferred from dust modeling for its 2002 and 2013 active apparitions, as well as expectations for sublimation-driven activity in general \citep{hsieh2004_133p,jewitt2014_133p}.
As mentioned in Section~\ref{section:H2O_analysis}, if we consider the nominal 25\% decline to be real, it is within the range of predicted amounts of decline in the water sublimation rate due to the increase in heliocentric distance between the two visits. However, the exact expected change due to heliocentric distance alone is dependent on 133P's rotational pole orientation, which is currently not definitively known \citep[see][]{kaluna2011_133pactivity}, and there are also several other factors that could have affected the water production rate evolution.

In the case of 67P, 
observations from the orbiting Rosetta spacecraft revealed widely varying terrain types across its surface \citep{thomas2015_67p}, each with different topographic, textural, and compositional properties, where those differences, along with spatially and temporally varying illumination conditions of different regions of the nucleus led to non-uniform temporal evolution of water production rates across the nucleus's surface \citep[e.g.,][]{marshall2017_67p,attree2023_67pactivitydistribution}, and so a similar situation could be the case for 133P as well.
Otherwise, even if 133P's surface is relatively uniform in terms of terrain type, its activity strength could also be attenuated (or otherwise modulated) by mantling \citep[e.g.,][]{jewitt1996_dormantcomets}, dust fall-back \citep[e.g.,][]{keller2017_67Pmasstransfer,marschall2020_67Pdust}, as well as constantly varying illumination and shadowing conditions of different portions (with different total surface areas) of its surface \citep[e.g.,][]{marschall2020_cometcomasurfacelinks}.

Of course, untangling these effects on 133P (or any other MBC) will likely only be possible via a Rosetta-like mission of its own \citep[e.g.,][]{meech2015_proteus,snodgrass2018_castalia,jones2018_caroline}.  In the meantime, efforts like finer temporal sampling of water production rates by \jwst{}, time-resolved ground-based reflectance spectroscopy of the nucleus to characterize surface inhomogeneity, and detailed shape modeling for improving thermal models could be useful for better characterizing water production rate variability and potential contributing factors to that variability.

\subsubsection{Dust Production}\label{section:dust_production_discussion}

Our inferred dust production rate during the first NIRSpec visit is consistent with the upper end of the range of dust production rates estimated for 133P by \citet{jewitt2014_133p} for its 2013 active apparition ($Q_d\sim(0.2-1.4)$~kg~s$^{-1}$), while our inferred dust production rate during the second NIRSpec visit is a factor of a few larger than the upper limit production rate estimate from the 2013 apparition.  However, those authors assumed $\rho_d=1000$~kg~m$^{-3}$, while we assume $\rho_d=2500$~kg~m$^{-3}$ here, in line with more recent MBC dust analyses \citep[e.g.,][]{hsieh2023_mbcnuclei,hsieh2025_358p}. We adopt this larger assumed grain density based on spectroscopic and dynamical evidence that 133P (and other MBCs) is compositionally more similar to denser C-complex asteroids than lighter classical comet nuclei from the outer solar system \citep[e.g.,][]{bagnulo2010_133p,licandro2011_133p176p,hsieh2018_activeastfamilies,depra2020_lixiaohua}, and should therefore produce similarly denser dust grains.  Since $M_d$ scales directly with $\rho_d$ (see Equation~\ref{eqn:dust_mass}), the dust production rate does as well, meaning that within available constraints, 133P's dust production rates in 2013 and 2024, as estimated by \citet{jewitt2014_133p} and this work, respectively, are similar, at least within an order of magnitude, for the same assumed grain density.

The estimated minimum aperture crossing time of $\sim70$~days for ejected dust particles computed above is much longer than the expected residence time of gas species produced at the same time \citep[which for our modeled H$_2$O velocity of $\sim500$~m~s$^{-1}$ at 2.3-2.6~au should be on the order of a few hours;][]{cochran1993h2o}.  As such, we expect a delay in the response of the measured excess photometric flux in the vicinity of the nucleus (expected to be dominated by dust) to changes in the dust production rate from the nucleus.  Inferred dust production rates from photometric proxy measurements like $Af\rho$ are therefore likely to be a lagging indicator of the actual dust production rate at the time of observation (particularly when larger and therefore slower dust particles are involved). This is the likely reason for the increase in the dust production rate of 133P inferred from optical photometric measurements between the two NIRSpec visits, despite the nominal decrease in water production rates measured from those NIRSpec observations over the same period. This apparent delayed response of photometric activity indicators like $Af\rho$ to changing activity is the same behavior implicated as being the likely explanation for observations of Jupiter-family comets (JFCs) largely systematically exhibiting maximal $Af\rho$ measurements during the post-perihelion portions of their orbits in data from ATLAS survey data \citep{gillan2024_atlasjfcs,gillan2025_atlastjfcs2}.

\subsubsection{Reflectance Spectroscopy}\label{section:refl_spectroscopy_discussion}

Detailed examination of 133P's reflectance spectra from NIRSpec reveals interesting similarities with other objects.  We first note that the 3~\textmu m absorption band in both epochs specifically resembles those of 67P/Churymov-Gerasimenko \citep[the target of the European Space Agency's Rosetta mission;][]{Taylor2017} and Cybele \citep{raponi2020_cometorganics,rivkin2022_3micron}, with its sharp edge at 2.7~\textmu m, maximum absorption at 3.1~\textmu m, and shallower redward slope.
There is also significant resemblance in the corrected 133P spectra to the spectra obtained by \citet{urso2020_organicirradiation} of samples of water ice, methanol, and ammonia, that were irradiated and then heated to 300 K. In particular, the shallow slope at 3.2$-$3.4~$\mu$m is quite similar to their Sample 2.

These results suggest that 133P may have formed beyond the ammonia ice line, making it similar to the other objects in the non-sharp type (NST) low-albedo asteroid population described by \citet{rivkin2022_3micron}. Comparing the reflectance spectrum of 133P to that paper's Figure 9, it is clear that the band shape of the NST average is similar in shape, but not in absorption depth. Both 133P datasets show a peak depth from 15\% to 20\%, significantly higher than the NST average \citep[see Table 10 of][]{rivkin2022_3micron}. The implications of this result are described further in Section \ref{sec:reflectance_spectroscopy_comparison}.  We do stress that our reduction and extraction techniques are optimized for gas sensitivity; therefore it is possible that we have included signal in the reflectance spectrum that is not representative of the nuclear flux, although we show that this is likely at the few percent level in Section \ref{section:refl_spectroscopy}.

\begin{figure*}[ht]
    \centering
    \includegraphics[width=0.8\linewidth]{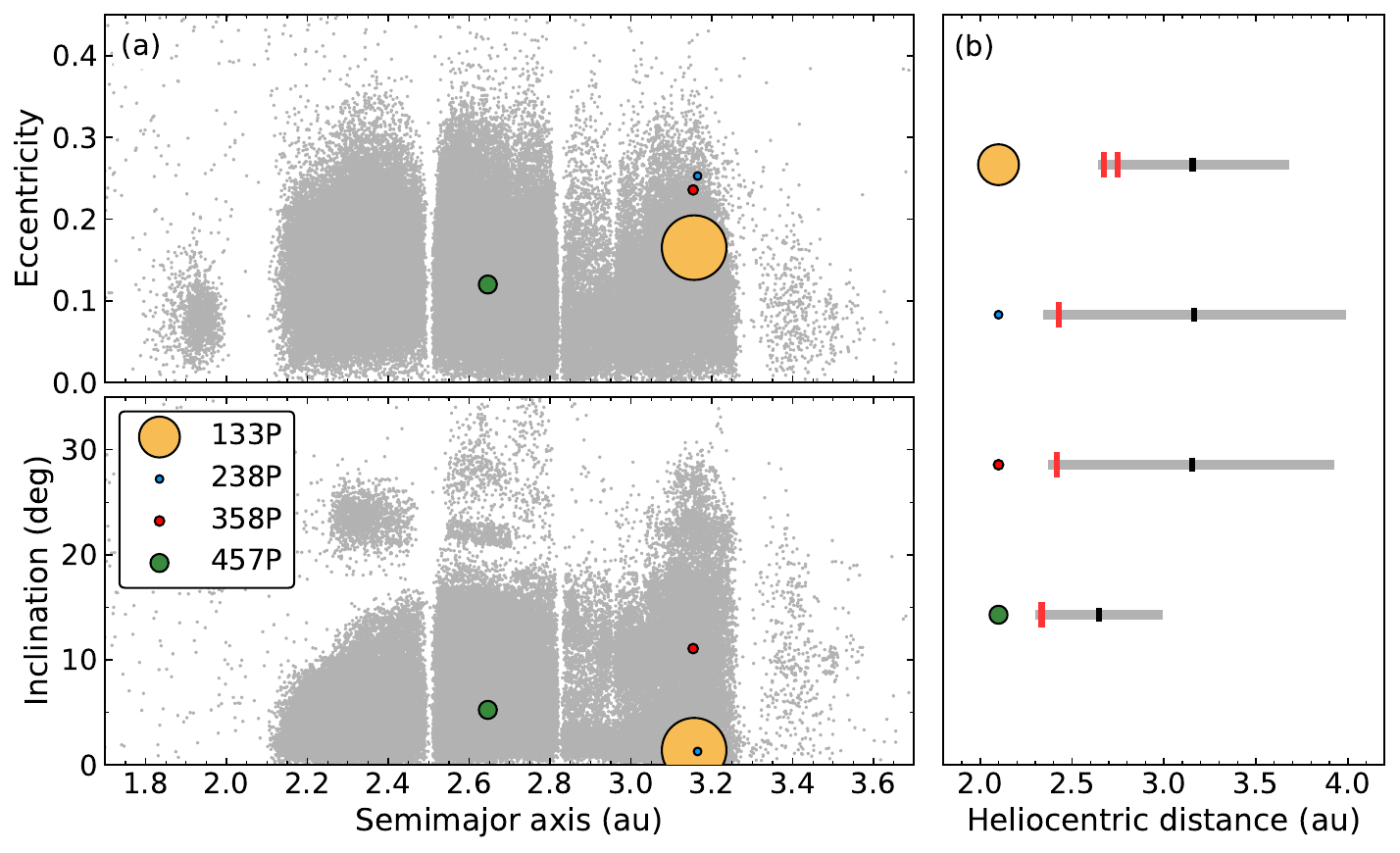}
    \caption{(a) Semimajor axis versus eccentricity (top panel) and inclination (bottom panel) plots showing known asteroids (small gray dots) and the four MBCs observed to date by \jwst{} (colored circular symbols, as labeled), based on values tabulated in Table~\ref{table:jwst_mbc_observations_summary}, where symbol sizes for MBCs are proportional to the measured sizes of each object's nucleus.  (b) Plot showing the heliocentric distances (red vertical tick marks) at which each MBC was observed by \jwst{}, where horizontal gray line segments extend from the perihelion to aphelion of each object's orbit, small vertical black tick marks show the semimajor axis of each object's orbit, and the same colored symobls representing individual MBCs in Panel (a) indicate which orbit corresponds to which object.  All targets observed to date were observed while on post-perihelion (i.e., outbound from the Sun) portions of their orbits.}
    \label{fig:aeiqQr}
\end{figure*}

\subsection{Comparison of Volatile and Dust Activity Parameters Among MBCs}\label{section:comparison}

\setlength{\tabcolsep}{5pt}
\setlength{\extrarowheight}{0em}
\begin{table*}[htb]
\caption{Main-Belt Comet Water Production Parameters$^a$}
\centering
\smallskip
\footnotesize
\begin{tabular}{ccccccc}
\hline\hline
\multicolumn{1}{c}{Target}
 & \multicolumn{1}{c}{Obs.\ Date$^b$}
 & \multicolumn{1}{c}{$Q_{\rm H_2O}$$^c$}
 & \multicolumn{1}{c}{$Q_{\rm H_2O}$$^d$}
 & \multicolumn{1}{c}{${\dot m_w}$$^e$}
 & \multicolumn{1}{c}{$A_{\rm act}$$^f$}
 & \multicolumn{1}{c}{$f_{\rm act}$$^g$}
 \\
\multicolumn{1}{c}{}
 & \multicolumn{1}{c}{}
 & \multicolumn{1}{c}{molecules~s$^{-1}$}
 & \multicolumn{1}{c}{kg~s$^{-1}$}
 & \multicolumn{1}{c}{}
 & \multicolumn{1}{c}{}
 & \multicolumn{1}{c}{}
 \\
\hline %
133P & 2024 Jun 12 & $(1.9\pm0.6)\times10^{25}$ & 0.6$\pm$0.2 &  0.4$-$2.4 & 8$-$49 & 0.0016$-$0.0097 \\ % obs=2460473.8 q=2460440.8
...  & 2024 Oct 14 & $(1.4\pm0.4)\times10^{25}$ & 0.4$\pm$0.1 & 0.3$-$2.2 & 6$-$48 & 0.0013$-$0.0096 \\ % obs=2460597.8 q=2460440.8
238P & 2022 Sep 08 & $(9.9\pm1.0)\times10^{24}$ & 0.30$\pm$0.03 & 0.9$-$3.4 &3$-$11 & 0.04$-$0.15  \\
358P & 2024 Jan 08 & $(5.0\pm0.2)\times10^{25}$ & 1.50$\pm$0.06 & 1.0$-$3.5 & 14 & 0.13 \\
457P & 2024 Sep 20 &  $<2.0\times10^{24}$ & $<$0.06 & 1.2$-$3.9  & $<$1.7 & $<$0.004  \\
\hline
\hline
\multicolumn{7}{l}{$^a$ References: 133P (this work); 238P \citep{kelley2023_jwst238p};} \\
\multicolumn{7}{l}{$~~~~$ 358P \citep{hsieh2025_358p}; 457P \citep{noonan2025_jwst457p}.} \\
\multicolumn{7}{l}{$^b$ UT observation date by NIRSpec on \jwst{}.} \\
\multicolumn{7}{l}{$^c$ Measured H$_2$O production rate in molecules~s$^{-1}$.} \\
\multicolumn{7}{l}{$^d$ Measured H$_2$O production rate in kg~s$^{-1}$.} \\
\multicolumn{7}{l}{$^e$ Range of predicted water sublimation rates in 10$^{20}$~kg~s$^{-1}$~m$^{-2}$.} \\
\multicolumn{7}{l}{$^f$ Inferred effective active area, in 10$^4$~m$^2$.} \\
\multicolumn{7}{l}{$^g$ Inferred effective active fraction.} \\
\end{tabular}
\label{table:mbc_water_production}
\end{table*}

\setlength{\tabcolsep}{7pt}
\setlength{\extrarowheight}{0em}
\begin{table*}[htb]
\caption{Main-Belt Comet Dust Activity Parameters$^a$}
\centering
\smallskip
\footnotesize
\begin{tabular}{cccccc}
\hline\hline
\multicolumn{1}{c}{Target}
 & \multicolumn{1}{c}{Obs.\ Date$^a$}
 & \multicolumn{1}{c}{$Q_d$$^d$}
 & \multicolumn{1}{c}{$Q_d/Q_{\rm H_2O}$$^e$}
 & \multicolumn{1}{c}{$Af\rho$$^f$}
 & \multicolumn{1}{c}{$\log{Af\rho / Q_{\rm H_2O}}$$^g$}
 \\
\hline %
133P & 2024 Jun 12 & 1.0$\pm$2.1 & $1.7\pm3.5$ & 0.8$\pm$1.5 & $~~$$-$25.4$\pm$0.8  \\ % obs=2460473.8 q=2460440.8
...  & 2024 Oct 14 & 3.8$\pm$2.1 & $9.5\pm5.8$ & 2.8$\pm$1.5 & $~~$$-$24.7$\pm$0.3  \\ % obs=2460597.8 q=2460440.8
238P & 2022 Sep 08 & 0.25 & 0.8 & 4.2$\pm$1.3 & $~~$$-$24.4$\pm$0.2   \\
358P & 2024 Jan 08 & 0.8 & 0.5 & 8.6$\pm$1.5 & $~~$$-$24.8$\pm$0.2  \\
457P & 2024 Sep 20 & 0.035$\pm$0.025 & $>$0.5 & 0.9$\pm$0.7 & $>$$-$24.4$\pm$0.3    \\
\hline
\hline
\multicolumn{6}{l}{$^a$ References: 133P (this work); 238P \citep{kelley2023_jwst238p};} \\
\multicolumn{6}{l}{$~~~~$ 358P \citep{hsieh2025_358p}; 457P \citep{noonan2025_jwst457p}.} \\
\multicolumn{6}{l}{$^b$ UT observation date by NIRSpec on \jwst{}.} \\
\multicolumn{6}{l}{$^c$ Estimated dust production rate in kg~s$^{-1}$.} \\
\multicolumn{6}{l}{$^d$ Estimated dust-to-water vapor production rate ratio by mass.} \\
\multicolumn{6}{l}{$^e$ Estimated $Af\rho$ value for a 5000~km photometry aperture at time of observation} \\
\multicolumn{6}{l}{$~~~~$ by NIRSpec, in cm.} \\
\multicolumn{6}{l}{$^f$ Dust-to-gas production rate ratios, as parameterized by the base-10 logarithm of} \\
\multicolumn{6}{l}{$~~~~$ the ratio between $Af\rho$ in cm and $Q_{\rm H_2O}$ in molecules~s$^{-1}$.} \\
\end{tabular}
\label{table:dust_vs_water}
\end{table*}

\begin{figure*}[ht]
    \centering
    \includegraphics[width=0.6\linewidth]{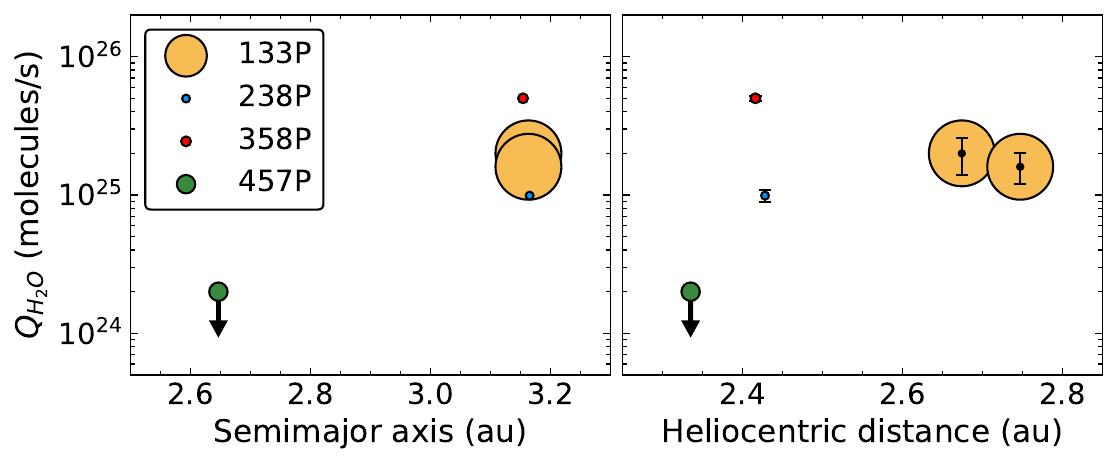}
    \caption{Measured $Q_{\rm H_2O}$ production rates for the five observations of four unique MBCs by \jwst{} to date (colored circular symbols, as labeled) plotted versus semimajor axis of each object (left panel) and heliocentric distance of the object at the time of observation by \jwst{} (right panel), where symbol sizes for MBCs are proportional to the measured sizes of each object's nucleus.  For reference, uncertainties are marked in the right panel, where they are smaller than the symbol sizes in the cases of 133P and 358P, and so in the case of 133P, are shown as part of additional markers (small black dots) overlaid on the primary 133P symbols. Uncertainties are not shown in the left panel for clarity (due to overlapping symbols), but are the same as in the right panel.}
    \label{fig:water_prod}
\end{figure*}

\begin{figure*}[ht]
    \centering
    \includegraphics[width=0.6\linewidth]{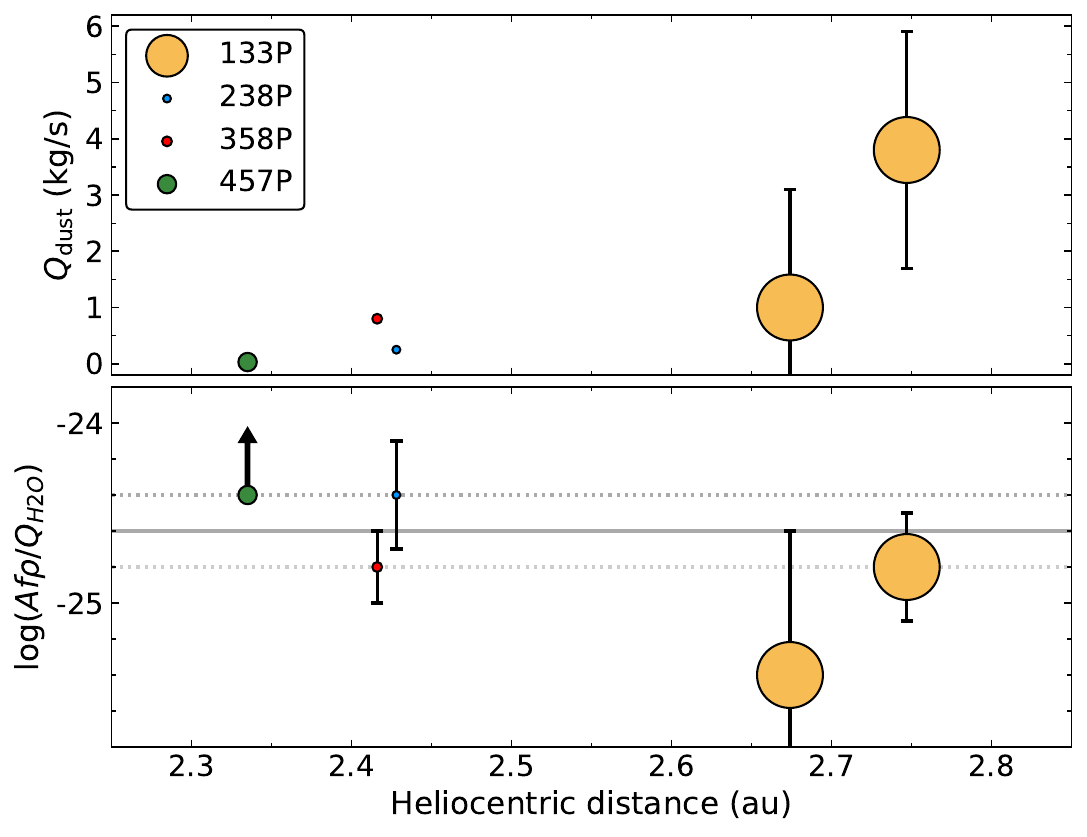}
    \caption{Inferred dust production rates (top panel) and measured $\log(Af\rho/Q_{\rm H_2O})$ values (bottom panel) for the four MBCs observed by \jwst{} to date (colored circular symbols, as labeled) plotted versus the heliocentric distance of the object at the time of observation by \jwst{}, based on values tabulated in Table~\ref{table:dust_vs_water}, where symbol sizes for MBCs are proportional to the measured sizes of each object's nucleus.  For reference, uncertainties are marked in both panels, where they are smaller than the symbol sizes in the cases of 238P, 358P, and 457P in the top panel.  In the bottom panel, the average $\log(Af\rho/Q_{\rm H_2O})$ for 238P, 358P, and the second visit to 133P computed in Section~\ref{section:dust_to_gas} is shown as a solid gray horizontal line, with the boundaries of the uncertainty range as computed from the sample's standard deviation shown as dotted gray horizontal lines.}
    \label{fig:dust_prod}
\end{figure*}

\setlength{\tabcolsep}{6pt}
\setlength{\extrarowheight}{0em}
\begin{table*}[htb]
\caption{Other Main-Belt Comet Volatile Activity Parameters}
\centering
\smallskip
\footnotesize
\begin{tabular}{ccccccccccc}
\hline\hline
\multicolumn{1}{c}{}
 & \multicolumn{1}{c}{}
 & \multicolumn{2}{c}{$Q_{\rm CO}$$^b$}
 & \multicolumn{2}{c}{$Q_{\rm CO_2}$$^c$}
 & \multicolumn{2}{c}{$Q_{\rm CH_3OH}$$^d$}
 & \multicolumn{1}{c}{$Q_{\rm CO}/$}
 & \multicolumn{1}{c}{$Q_{\rm CO_2}/$}
 & \multicolumn{1}{c}{$Q_{\rm CH_3OH}/$}
 \\
\multicolumn{1}{c}{Target}
 & \multicolumn{1}{c}{Obs.\ Date$^a$}
 & \multicolumn{1}{c}{molec~s$^{-1}$}
 & \multicolumn{1}{c}{kg~s$^{-1}$}
 & \multicolumn{1}{c}{molec~s$^{-1}$}
 & \multicolumn{1}{c}{kg~s$^{-1}$}
 & \multicolumn{1}{c}{molec~s$^{-1}$}
 & \multicolumn{1}{c}{kg~s$^{-1}$}
 & \multicolumn{1}{c}{$Q_{\rm H_2O}$$^e$}
 & \multicolumn{1}{c}{$Q_{\rm H_2O}$$^f$}
 & \multicolumn{1}{c}{$Q_{\rm H_2O}$$^g$}
 \\
\hline
133P & 2024 Jun 12 & $<$9.8$\times$10$^{24}$ & $<$0.5 & $<$1.4$\times$10$^{23}$ & $<$0.01 & $<$3.3$\times$10$^{23}$ & $<$0.02 & $<$0.5 & $<$0.007 & $<$0.02 \\ % obs=2460473.8 q=2460440.8
...  & 2024 Oct 14 & $<$5.5$\times$10$^{24}$ & $<$0.3 & $<$1.3$\times$10$^{23}$ & $<$0.01 & $<$4.9$\times$10$^{23}$ & $<$0.03 & $<$0.4 & $<$0.009 & $<$0.04 \\ % obs=2460597.8 q=2460440.8
238P & 2022 Sep 08 & --- & --- & $<$7$\times$10$^{22}$ & $<$0.005 & --- & --- & --- & $<$0.007 & --- \\
358P & 2024 Jan 08 & $<$3.0$\times$10$^{24}$ & $<$0.1 & $<$7.6$\times$10$^{22}$ & $<$0.006 & $<$6.4$\times$10$^{23}$ & $<$0.03 & $<$0.09 & $<$0.002 & $<$0.02 \\
457P & 2024 Sep 20 & $<$4.8$\times$10$^{24}$ & $<$0.2 & $<$1.4$\times$10$^{23}$ & $<$0.01 & $<$2.2$\times$10$^{24}$ & $<$0.1 & --- & --- & --- \\
\hline
\hline
\multicolumn{11}{l}{$^a$ References: 133P (this work); 238P \citep{kelley2023_jwst238p}; 358P \citep{hsieh2025_358p}; 457P \citep{noonan2025_jwst457p}.} \\
\multicolumn{11}{l}{$^a$ Observation date by NIRSpec on \jwst{}.} \\
\multicolumn{11}{l}{$^b$ Measured upper limit CO production rate in molecules~s$^{-1}$ and kg~s$^{-1}$.} \\
\multicolumn{11}{l}{$^c$ Measured upper limit CO$_2$ production rate in molecules~s$^{-1}$ and kg~s$^{-1}$.} \\
\multicolumn{11}{l}{$^d$ Measured upper limit CH$_3$OH production rate in molecules~s$^{-1}$ and kg~s$^{-1}$.} \\
\multicolumn{11}{l}{$^e$ CO to H$_2$O production rate ratio (by number of molecules).} \\
\multicolumn{11}{l}{$^f$ CO$_2$ to H$_2$O production rate ratio  (by number of molecules).} \\
\multicolumn{11}{l}{$^g$ CH$_3$OH to H$_2$O production rate ratio  (by number of molecules).} \\
\multicolumn{11}{l}{$^h$ References: [1] This work; [2] \citet{kelley2023_jwst238p}; [3] \citet{hsieh2025_358p}; [4] \citet{noonan2025_jwst457p}.} \\
\end{tabular}
\label{table:mbc_other_volatiles}
\end{table*}

\subsubsection{Overview\label{section:comparison_overview}}

To date, \jwst{} observations have been obtained for four unique MBCs: 238P \citep{kelley2023_jwst238p}, 358P \citep{hsieh2025_358p}, 133P (this work), and 457P \citep{noonan2025_jwst457p}, where 133P was observed twice.  We present a summary of the findings in those studies in Tables~\ref{table:jwst_mbc_observations_summary}, \ref{table:mbc_water_production}, \ref{table:dust_vs_water}, and \ref{table:mbc_other_volatiles}. 
Table~\ref{table:jwst_mbc_observations_summary} lists orbital and physical parameters of the MBCs observed to date by \jwst{} and the circumstances of those observations.
Table~\ref{table:mbc_water_production} lists measured water production rates or limits and inferred active areas and surface fractions.
Table~\ref{table:dust_vs_water} lists various measured and inferred dust-related activity parameters.
Lastly, Table~\ref{table:mbc_other_volatiles} lists upper limit production rates and corresponding volatile ratio limits determined for various cometary volatile species other than water.  The large size of the potential available parameter space sampled by these observations (including but not limited to reasonably well-known parameters like object sizes, orbital elements, and orbit positions, and poorly constrained parameters like size-frequency distribution of ejected dust particles, rotational properties, detailed compositional and structural nucleus characteristics, surface terrain, formation circumstances, and thermal and activity history, to name just a few) means that it is impossible to draw any firm conclusions from the limited currently available data.  However, we can still make a number of interesting observations from our current body of measurements.

\subsubsection{Water Production Rates\label{section:comparison_waterproduction}}

When assessing the present state of \jwst{} detections of water vapor outgassing from MBCs, we first note that they are generally consistent with thermal evolution studies, with models showing that water ice could persist within meters or less of the surface of an outer main-belt object like 133P over Gyr timescales \citep{schorghofer2008_mbaice,schorghofer2016_asteroidice,schorghofer2018_asteroidiceloss}. Water ice at these depths would be well within the reach of various activity triggering mechanisms like small impacts or rotationally triggered mass-wasting events \citep[e.g.,][]{haghighipour2016_mbcimpacts,haghighipour2018_mbcactivation,hsieh2018_activeastfamilies,hsieh2023_mbcnuclei}.  \jwst{}'s detections of outgassing are also in agreement with the long-argued assertion that sublimation is the only plausible mechanism for explaining the particular characteristics of MBC activity, namely long durations for dust emission events and recurrence near perihelion with intervening periods of inactivity (see Section~\ref{section:background}).

In terms of specific findings to date, we find that water production rates for the three MBCs for which these rates have been successfully measured are similar to within an order of magnitude, with the water production rate limit for the fourth MBC being about an order of magnitude lower (Figure~\ref{fig:water_prod}).  
These water production rates do not appear to be simply correlated with a single physical, orbital, or observational parameter.

For example, 133P and 238P exhibit very similar water production rates (see Table~\ref{table:mbc_water_production}) despite being the largest and smallest objects, respectively.  Meanwhile, 238P and 358P exhibit water production rates that vary by a factor of 5$\times$ despite having very similar semimajor axes and being observed at very similar heliocentric distances (see Table~\ref{table:jwst_mbc_observations_summary}), and the objects with the largest and smallest water production rates (358P and 457P) are the two objects at the smallest heliocentric distances at the time of observation.

Of potential interest, the MBC in our current sample with the largest measured water production rate (358P) also has the largest inclination, while the MBC with the smallest measured water production rate upper limit (457P) has the smallest semimajor axis, and is also the only MBC observed to date with a semimajor axis interior to the 5A:2J mean-motion resonance (MMR) with Jupiter at $a=2.824$~au. 
In the former case, 358P's higher water production rate could be attributable to the fact that collisions are less frequent in the main asteroid belt at larger inclinations \citep[e.g.,][]{farinella1992_astcollisionrates}.  Assuming that collisions are a primary activity triggering mechanism for MBCs \citep[e.g.,][]{haghighipour2016_mbcimpacts,hsieh2018_activeastfamilies}, 358P might therefore have experienced fewer triggering collisions over its lifetime, and therefore has retained more ice that can support stronger emission when a triggering impact does occur.  We note that this is consistent with 324P/La Sagra, another high-inclination MBC ($i=21.4^{\circ}$), which has also been shown to have stronger activity (relative to its nucleus size) than other MBCs with lower inclinations \citep{hsieh2012_324p}.

In the latter case, 457P's weak activity could be the result of the more rapid depletion of its ice reservoirs relative to other MBCs due to its closer proximity to the Sun, in analogy to conclusions reached for 259P/Garradd \citep{hsieh2021_259p}, which like 457P, also has a semimajor axis interior to the 5A:2J mean-motion resonance with Jupiter.  This reasoning is also consistent with the much smaller heliocentric distance of 457P's activity onset point \citep[$r_h\sim2.35$~au;][]{noonan2025_jwst457p} compared to those for MBCs with semimajor axes exterior to the 5A:2J mean-motion resonance with Jupiter
\citep[e.g.,][]{hsieh2018_358p,hsieh2018_238p288p}, 
which in 259P's case, was attributed to ice reservoirs being located at greater depths than on other MBCs due to the increased ice depletion rate.

Overall, five sets of \jwst{} observations of four unique objects is of course far too small a sample to discern the effects of the numerous parameters that could affect MBC water production rates, but it is still useful to assess whether any early trends or early points of potential interest are emerging in order to better guide future observing strategies.  To this point, while continued characterization of the volatile activity of any MBCs by \jwst{} will be useful in general, we suggest that \jwst{} observations of more MBCs interior to the 5A:2J MMR with Jupiter would be particularly useful for determining if 457P's extremely low water production rate is typical or anomalous for MBCs in this region of the asteroid belt.
Observing more high-inclination MBCs like 358P would show whether they exhibit consistently higher water production rates than lower-inclination MBCs.
Lastly, observing individual MBCs multiple times during a single active apparition, as was done for 133P, would enable evaluation of how water production rates evolve for other objects over the course of a single apparition.

\subsubsection{Active Surface Fractions}\label{section:active_fractions}

We find a wide range of active surface fractions ($f_{\rm act}$) among the currently observed MBCs with successfully detected water vapor outgassing, with the smallest and largest potential active surface fractions (depending on the spin states assumed for each objects) differing by about two orders of magnitude (see Table~\ref{table:mbc_water_production}).  For context, the estimated active fractions for 238P and 358P are well within the range of active fractions found for classical comets \citep[i.e., a few percent or more;][]{ahearn1995_ensemblecomets,fernandez1999_jfcpopulation,samarasinha2013_c2012s1}.  Meanwhile, the corresponding active fractions for 133P and 457P are potentially about an order of magnitude smaller than the vast majority of classical comets, but may still have analogs in extremely low-activity JFCs with $f_{\rm act}<1$\% \citep[e.g., 10P/Tempel 2;][]{samarasinha2013_c2012s1,schleicher2013_tempel2}.

Interestingly, the significantly smaller active surface fraction inferred for 133P relative to 238P and 358P is offset by 133P's significantly larger nucleus size (see Table~\ref{table:jwst_mbc_observations_summary}), leading to absolute active surface areas being consistent within about an order of magnitude for all three objects (see Table~\ref{table:mbc_water_production}).  This result is notably similar to those of \citet{hsieh2014_324p}, who similarly found that inferred active surface areas (derived from dust mass loss rates computed from numerical dust modeling and an assumed dust-to-gas ratio of $f_{dg}=10$) for a sample of eight MBCs (including 133P, 238P, and 358P) were similar to within two orders of magnitude, despite a much wider range in inferred active surface fractions.  \citet{hsieh2014_324p} suggested that this result could potentially be explained by the active areas on these MBCs all being excavated by impactors of similar size (assuming that impact-activation was responsible for activity on every object).  Further speculation on this subject is beyond the scope of this work, and as such, we simply note for now that these results may be useful to consider in future studies considering potential MBC activation mechanisms.

\subsubsection{Dust to Gas Ratios\label{section:comparison_dusttogas}}

As discussed in Section~\ref{section:dust_to_gas}, dust-to-gas ratios as parameterized by $\log(Af\rho/Q_{\rm H_2O})$ for the three MBCs for which water vapor outgassing has been successfully detected, are consistent within uncertainties with an average value of $\log(Af\rho/Q_{\rm H_2O})=-24.6\pm0.2$ (see Table~\ref{table:dust_vs_water}), provided that an additional data point associated with 133P is excluded due to large uncertainties.  This is notable in that it could open the possibility of estimating (and therefore comparing) water production rates over many more MBCs than those individually observed by \jwst{}, as well as for archival observations of earlier active epochs of MBCs obtained prior to \jwst{}'s launch and future data obtained after the end of its operational lifetime.

The details of these results are instructive, however, in that they show that, despite the consistency of most of the measured values of $\log(Af\rho/Q_{\rm H_2O})$ among our sample of MBC observations, outlier values are possible. In the case of this work, the outlying value was computed at a time when $Af\rho$ had a particularly large uncertainty with respect to its nominal value (Section~\ref{section:afrho_optical}) leading to an anomalously large uncertainty for $\log(Af\rho/Q_{\rm H_2O})$.  During this time, even though we were able to achieve a well-constrained measurement of $Q_{\rm H_2O}$ using NIRSpec, the dust coma had not yet become bright enough to overcome periodic negative brightness ``enhancements'' due to 133P's rotational lightcurve in optical observations.  As such, it was not always possible to formally compute $Af\rho$ for the dust for individual snapshot observations (i.e., observations that did not capture the full rotational lightcurve that would enable accurate estimation of the photometric midpoint), leading us to pursue alternate means for estimating an average value for the entire period (see Section~\ref{section:afrho_optical}).  Meanwhile, the $\log(Af\rho/Q_{\rm H_2O})$ measured for 133P that was more consistent with previous measurements for 238P and 358P was acquired at a time when the average value of $Af\rho$ was much better constrained.

These results suggest that even assuming dust-to-gas production rate ratios are relatively stable over the course of a MBC's active apparition, the ability to measure $Af\rho$ accurately and precisely is also (unsurprisingly) extremely important for inferring reliable water production rates from an assumed standard value of $\log(Af\rho/Q_{\rm H_2O})$.  Based on the work presented here, we would suggest that water production rates estimated from any $Af\rho$ measurements with uncertainties more than 50\% of nominal values be regarded with significant caution.  
In the specific case of 133P, it would have been useful to have acquired full lightcurves around the time of our first NIRSpec visit, so that its photometric midpoint at the time could have been more accurately estimated. As such, we would recommend that approach for any future observations of MBCs with known large rotational lightcurve variations at times when dust production is expected to be extremely weak (e.g., near the start of activity).

Further \jwst{} observations for other MBCs will certainly be required, however, to determine whether the average value of $\log(Af\rho/Q_{\rm H_2O})=-24.6\pm0.2$ found for our current MBC sample continues to be broadly consistent.
It will also be useful to determine if there is a particular range of orbit positions over which that average value applies (e.g., not extremely early or extremely late in their active apparitions when water production are more likely to be decoupled from dust production rate proxy measurements like $Af\rho$; e.g., Section~\ref{section:dust_production_discussion}).

Notably, despite the relative consistency between $\log(Af\rho/Q_{\rm H_2O})$ values computed for 133P, 238P, and 358P, the dust-to-gas production rate ratio by mass of $Q_d/Q_{\rm H_2O}=9.5\pm5.8$ calculated for that second visit of 133P (Section~\ref{section:dust_to_gas}) is significantly larger than the corresponding ratios for 238P and 358P (Table~\ref{table:dust_vs_water}).
As discussed in Section~\ref{section:dust_production_discussion}, however, dust production rate estimates are highly parameter dependent, and can vary by several factors depending on estimates or assumptions of particle size distributions, densities, and velocities (which can vary widely between different objects, between different active apparitions of the same object, and perhaps even within a single active apparition of a given object).  As such, we opt to refrain for now from ascribing too much meaning from comparisons across multiple objects of parameter values derived from these estimates.

\subsubsection{Hypervolatile Depletion and Correspondence to Thermal Modeling}\label{section:comparison_hypervolatiles}

The CO$_2$ to H$_2$O production rate ratio limits we find for 133P ($Q_{\rm CO_2}/Q_{\rm H_2O}<0.009$) are consistent with those measured for 238P and 358P \citep[$Q_{\rm CO_2}/Q_{\rm H_2O}<0.007$ and $Q_{\rm CO_2}/Q_{\rm H_2O}<0.002$, respectively;][]{kelley2023_jwst238p,hsieh2025_358p}, all of which are an order of magnitude lower than measurements of other comets at similar heliocentric distances.

No volatile species other than water have been observed for any MBCs observed to date, although meaningful constraints have only  been obtained for 133P, 238P, and 358P (see Section~\ref{section:minor_species_analysis}).  The upper limits on hypervolatile production rates (especially for CO$_2$) found for those objects are consistent with predictions from thermal modeling of strong depletion of hypervolatile species (specifically HCN, NH$_3$, CO$_2$, and C$_2$H$_2$) in main-belt objects over timescales of $\gg10^7$~yr, assuming they were accreted in appreciable amounts in the first place \citep{prialnik2009_mbaice}.
\jwst{} results to date therefore support conclusions from numerical integration studies that MBCs have had long residence times in the asteroid belt \citep[$\sim10^8$~yr timescales or longer; e.g.,][]{haghighipour2009_mbcorigins,hsieh2012_288p,hsieh2012_324p}.

A dynamical integration study by \citet{hsieh2016_tisserand}, however, suggested that some JFCs could temporarily evolve onto main-belt-like orbits (specifically ones with high eccentricities, $e$, and high inclinations, $i$), implying that some currently active MBCs (such as 358P) could actually be recently implanted JFC interlopers that may not exhibit the same degree of hypervolatile depletion as MBCs that were either formed in situ or delivered to the asteroid belt at much earlier times \citep[such as during the era of planetary migration; e.g.,][]{walsh2011_grandtack}  The strong CO$_2$ depletion found for 358P \citep{hsieh2025_358p} does not contradict this hypothesis, as the hypothesis merely predicts that some, but not all, high-$e$, high-$i$ MBCs are recently implanted interlopers.  That said, in light of that dynamical work, additional \jwst{} observations (at sufficiently high signal-to-noise that depletion measurements are meaningful) would be very useful for investigating whether other high-$e$, high-$i$ MBCs show more JFC-like CO$_2$ abundances, implying recent arrival times to the asteroid belt.

\subsection{Reflectance Spectroscopy}
\label{sec:reflectance_spectroscopy_comparison}

A comparison of the reflectance spectra acquired from the JWST MBC observations thus far shows a broad diversity of results. \citet{noonan2025_jwst457p} showed that the reflectance spectrum of 457P appears to have more rounded 3 $\mu$m feature than 133P, with a band center nearer to 2.9~$\mu$m, a deeper 3.4~$\mu$m feature, and a shallower 2.7~$\mu$m-absorption slope, while both 238P \citep{kelley2023_jwst238p} and 358P \citep{hsieh2025_358p} appear to be more like 133P, but have substantially more dust that obscures our interpretation. What is clear is that if 457P is confirmed to have a substantially different surface composition while exhibiting likely sublimation-driven activity, albeit at a much lower level than the other three, a key question is whether the difference is intrinsic or evolutionary. Untangling that question will require a more specific examination of the reflectance spectra for each of the targets thus far, and more targeted observations of MBC nuclei at aphelion when activity is low and the contribution from dust even less significant to minimize sources of uncertainty. A search for changes to the absorption features over multiple orbits would also be informative to understand if newer, potentially less irradiated, material is being exposed, or if the active regions are localized and not effective in shedding mass on a global scale. 

\section{Summary and Conclusions}

In this work, we present the following key findings:
\begin{enumerate}
\item{We report water vapor outgassing rates at two different times for 133P/Elst-Pizarro, finding $Q_{\rm H_2O}=(1.9\pm0.6)\times10^{25}$~molecules~s$^{-1}$  on UT 2024 June 12 when the object was at a true anomaly of $\nu=8^{\circ}$ and a heliocentric distance of $r_h=2.674$~au, and $Q_{\rm H_2O}=(1.4\pm0.4)\times10^{25}$~molecules~s$^{-1}$  on UT 2024 October 14 when it was at $\nu=37.4^{\circ}$ and $r_h=2.747$~au.
These measurements correspond to a nominal 25\% decline in $Q_{\rm H_2O}$ between the two visits, which is within the range of predicted amounts of decline in the water sublimation rate due to the increase in heliocentric distance between the two visits.  However, within uncertainties, these measurements are also consistent with no change in $Q_{\rm H_2O}$ between the visits. Importantly, these measurements show the presence of detectable water vapor outgassing from 133P during observations four months apart, confirming the prolonged nature of the object's activity.}
\item{We find no spectroscopic evidence of CO, CO$_2$, or CH$_3$OH in either set of observations of 133P, with upper limits of $9.8\times10^{24}$~molecules~s$^{-1}$, $1.4\times10^{23}$~molecules~s$^{-1}$, and $4.9\times10^{23}$~molecules~s$^{-1}$, respectively. These upper limits place the comet's hypervolatile depletion at a similar level as previously observed MBCs, where we specifically find $Q_{\rm CO_2}/Q_{\rm H_2O}<0.009$ for 133P, compared to $Q_{\rm CO_2}/Q_{\rm H_2O}<0.007$ for 238P and $Q_{\rm CO_2}/Q_{\rm H_2O}<0.002$ for 358P.}
\item{We find 3~$\mu$m absorption bands in spectra from both visits to 133P which exhibit a significant amount of structure, and qualitatively resemble those of comet 67P/Churyumov-Gerasimenko and asteroid (65) Cybele. These results suggest that 133P could have formed beyond the ammonia ice line, similar to other objects in the non-sharp type population described by \citet{rivkin2022_3micron}.}
\item{NIRCam imaging of 133P on UT 2024 October 14 and 28 usng the F200W and F277W broadband filters show a point-source-like nucleus that deviates from circular symmetry beyond $\rho\sim0\farcs2$ from the photocenter in both filters during both visits.  Additionally, both sets of observations show a visible but low-brightness narrow linear tail with a position angle aligned with that of the anti-Solar vector, along with a much fainter Sun-ward tail in the approximate direction of the negative heliocentric velocity vector that is more visible in some contemporary ground-based optical observations.  There are no significant changes in this morphology between the two visits.}
\item{From NIRCam data, we measure average AB magnitudes of $m_{\rm F200W}=(20.98\pm0.14)$~mag and $m_{\rm F277W}=(21.49\pm0.11)$~mag on UT 2024 October 14, and average magnitudes of $m_{\rm F200W}=(21.38\pm0.08)$~mag and $m_{\rm F277W}=(21.91\pm0.10)$~mag on UT 2024 October 28, where the listed uncertainties correspond to the standard deviation of the individual photometric measurements included in each average.  These results correspond to ${\rm F200W}-{\rm F277W}$ colors of $-0.51\pm0.18$ on UT 2024 October 14 and $-0.53\pm0.13$ on UT 2024 October 28.  
}
\item{The morphology evolution of 133P from ground-based optical data is characteristic of an object displaying sublimation-driven activity, where a single antisolar tail is observed until UT 2024 August 27 when the object begins clearly displaying two tails, one in the antisolar direction, and one in the sunward direction aligned with the negative heliocentric velocity vector, a characteristic morphology indicating the simultaneous presence of older, larger particles and newer, smaller particles, and therefore implying a prolonged emission event consistent with sublimation.}
\item{We find extremely small $Af\rho$ values derived from ground-based optical observations of $Af\rho=(0.8\pm1.5)$~cm and $Af\rho=(2.8\pm1.5)$~cm at the times of our two NIRSpec observations. From these $Af\rho$ values, effective mean particle sizes determined from a previously published dust modeling analysis, and assumed particle densities, we infer dust production rates of $Q_d=(1.0\pm2.1)$~kg~s$^{-1}$ and $Q_d=(3.8\pm2.1)$~kg~s$^{-1}$ during our first and second NIRSpec observations, respectively.}
\item{ We find similar $\log(Af\rho/Q_{\rm H_2O})$ values for 133P (during one of our visits), 238P, and 358P, that are all consistent within uncertainties with an average value of $\log(Af\rho/Q_{\rm H_2O})=-24.6\pm0.2$.  This result is notable in that it opens the possibility of estimating water production rates for many more MBCs than those individually observed by \jwst{}, by leveraging the much larger body of optical photometric data from ground-based observations of MBCs, although additional \jwst{} observations of MBCs are encouraged to clarify the circumstances under which this result is valid.}
\item{In terms of the ensemble properties of the MBCs observed to date by \jwst{}, we so far find no straightforward dependencies of water production rates on parameters such as nucleus size, semimajor axis, or heliocentric distance at the time of observation.  We also particularly encourage continued JWST observations of additional MBCs interior to the 5A:2J MMR with Jupiter and at high inclinations, as well as observations of MBCs at multiple times during the same active apparitions to further investigate areas of interest that have been identified from the current sample of JWST MBC observations acquired thus far.
}
\end{enumerate}

\section*{Data and Software Availability\label{section:software}}

Pipeline-processed \jwst{} data are publicly available from the Space Telescope Science Institute's Mikulski Archive for Space Telescopes at \url{https://mast.stsci.edu/} under JWST program IDs 4250 and 5551, and at
\dataset[https://doi.org/10.17909/knbd-r965]{https://doi.org/10.17909/knbd-r965}.
Ground-based image data from Gemini Observatory are publicly available online from the Gemini Observatory Archive at \url{https://archive.gemini.edu/} under program IDs GS-2023A-LP-104, GS-2024A-Q-111, GN-2024B-Q-114, and GS-2024B-Q-113.

This work makes use of the Planetary Spectrum Generator \citep{villanueva2018_psg} at \url{https://psg.gsfc.nasa.gov/}, the Ice Sublimation Model at \url{https://github.com/Small-Bodies-Node/ice-sublimation}, and the \jwst{} science data calibration pipeline at \url{https://github.com/spacetelescope/jwst/}.
This research also makes use of the Jet Propulsion Laboratory's Horizons online ephemeris generation tool \citep{giorgini1996_horizons};
NASA's Astrophysics Data System Bibliographic Services (\url{https://ui.adsabs.harvard.edu/}), which is funded by NASA under Cooperative Agreement 80NSSC21M00561;
the Cometary Coma Image Enhancement Facility (\url{http://cie.psi.edu});
{\tt astropy}, a community-developed core {\tt python} package for astronomy;
{\tt ccdproc}, an {\tt astropy} package for image reduction;
{\tt L.A.Cosmic}, a cosmic ray rejection algorithm \citep{vandokkum2001_lacosmic};
{\tt pyraf}, a product of the Space Telescope Science Institute, which is operated by AURA for NASA;
{\tt sbpy}, an {\tt astropy} affiliated package for small-body planetary astronomy \citep{mommert2019_sbpy};
{\tt scipy}, an open-source software package for mathematics, science, and engineering \citep{virtanen2020_scipy};
a {\tt python} implementation of the \citet{cowan1979_cometsublimation} sublimation model \citep{vanselous21_ice_e20745b}; and
{\tt uncertainties} (version 3.0.2), a {\tt python} package for calculations with uncertainties by E.~O.\ Lebigot\footnote{\url{http://pythonhosted.org/uncertainties/}}.

\begin{acknowledgments}

We thank two anonymous reviewers for helpful feedback that improved this manuscript.
This work benefited from support from the International Space Science Institute, Bern, Switzerland, through the hosting and provision of financial support for an international team to discuss the science of main-belt comets.  This work is based on observations made with the NASA/ESA/CSA James Webb Space Telescope.  The data were obtained from the Mikulski Archive for Space Telescopes at the Space Telescope Science Institute, which is operated by the Association of Universities for Research in Astronomy, Inc., under NASA contract NAS 5-03127 for \jwst{}.  These observations are associated with \jwst{} General Observer Programs 4250 and 5551.
Support for this work was provided to H.H.H., J.W.N., M.S.P.K., and D.B.\ by NASA through grant NAS 5-03127  from the Space Telescope Science Institute, which is operated by the Association of Universities for Research in Astronomy, Inc., under contract NAS 5-26555.

H.H.H., M.S.P.K., J.P., S.S.S., and A.T.\ also acknowledge support from the NASA Solar System Observations program (Grant 80NSSC19K0869).  The work of J.P.\ was conducted at the Jet Propulsion Laboratory, California Institute of Technology, under a contract with the National Aeronautics and Space Administration (80NM0018D0004).
C.O.C. acknowledges support from the NASA CSSFP (grant No. 80NSSC26K0380), and Arthur and Jeanie Chandler. LINCC Frameworks is supported by Schmidt Sciences. C.O.C. also acknowledges support from the DiRAC Institute in the Department of Astronomy at the University of Washington. The DiRAC Institute is supported through generous gifts from the Charles and Lisa Simonyi Fund for Arts and Sciences.

The authors thank M.\ M.\ Knight for assistance in obtaining NTT observations,
N.\ Samarasinha for assistance in running Cometary Coma Image Enhancement Facility code, and C.\ Soto, B.\ Hilbert, K.\ Glidic, J.\ Stansberry, S.\ Meyett and other Space Telescope Science Institute staff for their assistance in obtaining \jwst{} observations.

This work is based on observations obtained at the international Gemini Observatory (under Programs GS-2023A-LP-104, GS-2024A-Q-111, GN-2024B-Q-114, and GS-2024B-Q-113), a program of NSF NOIRLab, which is managed by the Association of Universities for Research in Astronomy (AURA) under a cooperative agreement with the U.S. National Science Foundation on behalf of the Gemini Observatory partnership: the U.S. National Science Foundation (United States), National Research Council (Canada), Agencia Nacional de Investigaci\'{o}n y Desarrollo (Chile), Ministerio de Ciencia, Tecnolog\'{i}a e Innovaci\'{o}n (Argentina), Minist\'{e}rio da Ci\^{e}ncia, Tecnologia, Inova\c{c}\~{o}es e Comunica\c{c}\~{o}es (Brazil), and Korea Astronomy and Space Science Institute (Republic of Korea).
Portions of this work were specifically enabled by observations made from the Gemini North telescope, located within the Maunakea Science Reserve and adjacent to the summit of Maunakea. We are grateful for the privilege of observing the Universe from a place that is unique in both its astronomical quality and its cultural significance.

This work is also based on observations obtained at the Hale Telescope at Palomar Observatory as part of a continuing collaboration between the California Institute of Technology, NASA/JPL, Yale University, and the National Astronomical Observatories of China.

This work is also based on observations obtained at the Lowell Discovery Telescope at Lowell Observatory.
Lowell is a private, non-profit institution dedicated to astrophysical research and public appreciation of astronomy and operates the LDT in partnership with Boston University, the University of Maryland, the University of Toledo, Northern Arizona University and Yale University.
The Large Monolithic Imager was built by Lowell Observatory using funds provided by the National Science Foundation (AST-1005313). The University of Maryland LDT observing team consists of Q.\ Ye, J.\ Bauer, A.\ Gicquel-Brodtke, T.\ Farnham, L.\ Farnham, C.\ Holt, M.\ S.\ P.\ Kelley, J.\ Kloos, and J.\ Sunshine.

This work is also based on observations collected at the European Organisation for Astronomical Research in the Southern Hemisphere under ESO programme 113.26J9.002.

The authors thank 
T.\ Barlow, K.\ Koviak, P.\ Nied, and other Palomar Observatory staff;
NTT observatory staff; and
L.\ Alamos, J.\ Andrews, J.\ Ball, J.\ Berghuis, P.\ Candia, R. Carrasco, J.\ Chavez, K. Chiboucas, A.\ Cikota, B.\ Cooper, E.\ Deibert, E.~P.\ Farina, C.\ Figura, J.\ Font-Serra, J.\ Fuentes, J.-E.\ Heo, V.\ Kalari, E.\ Kim, S.\ Leggett, A.\ Lopez, C.\ Mart\'inez-V\'azquez, D.\ May, B.\ Miller, J.\ Miller, T.\ Mo{\v c}nik, P.~M.\ Ravelo, M.\ Rawlings, H.\ Reggiani, R.\ Ruiz, T.\ Seccull, K.\ Silva, A.\ Smith, A.\ Stephens, S.\ Stewart, J.\ Thomas-Osip, J.\ Turner, and other Gemini Observatory staff
for their assistance in obtaining observations.

\end{acknowledgments}

\renewcommand{\thesubsection}{\Alph{subsection}}

\appendix
\setcounter{figure}{0}
\renewcommand{\thefigure}{A\arabic{figure}}

\section{Composite Images from Ground-based Observations\label{section:appendix_groundbased_images}}

\begin{figure*}[ht]
    \centering
    \includegraphics[width=0.8\linewidth]{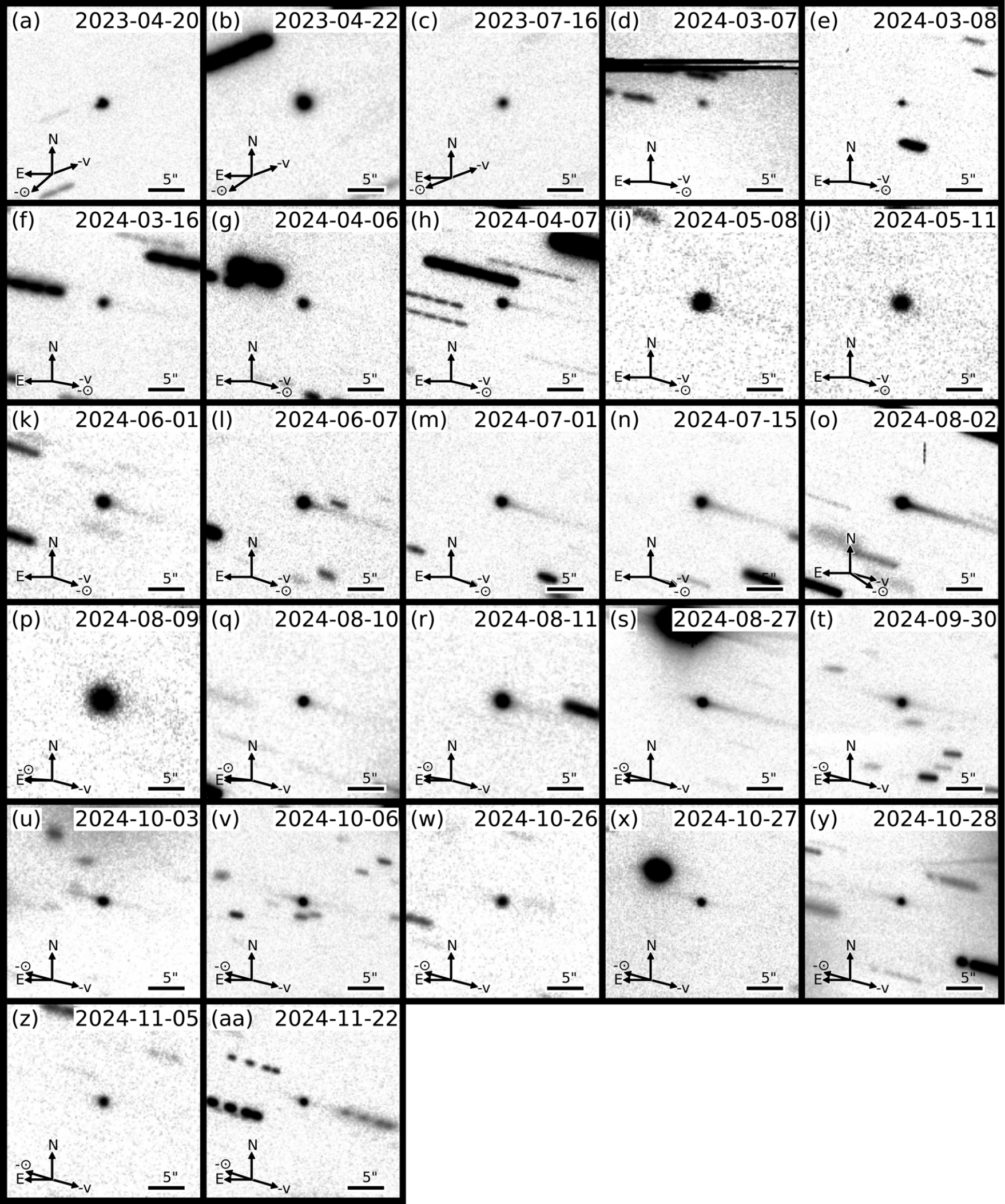}
    \caption{Composite images constructed from ground-based optical observations of 133P listed in Table~\ref{table:ground_observations_133p}.  Each panel includes the date of observation, a $5''$ scale bar to indicate the size of each image, and arrows denoting the directions of North (N), East (E), the antisolar vector projected on the sky ($-\odot$), and the negative heliocentric velocity vector projected on the sky ($-v$).
    \label{fig:133p_optical_images}}
\end{figure*}

\clearpage

\bibliography{main}{}
\bibliographystyle{aasjournalv7}

\end{CJK*}
\end{document}